\documentclass{mn2e}
\usepackage{times}
\usepackage{epsfig}
\def\spose#1{\hbox to 0pt{#1\hss}}
\def\oo{[O\textsc{ii}]}
\def\s2{[S\textsc{ii}]}
\def\o3{[O\textsc{iii}]}
\def\ne3{[Ne\textsc{iii}]}
\def\lta{\mathrel{\spose{\lower 3pt\hbox{$\mathchar"218$}}\raise 2.0pt\hbox{$\ma
thchar"13C$}}}
\def\gta{\mathrel{\spose{\lower 3pt\hbox{$\mathchar"218$}}\raise 2.0pt\hbox{$\ma
thchar"13E$}}}

\title[9C J1503+4528: a young radio source]{Deep spectroscopy of 9C J1503+4528: a very young CSS radio source at z=0.521}

\author[K.\,J.\, Inskip et al]
{K.\,J.\, Inskip$^{1}$\footnotemark, D.\, Lee$^{2}$, Garret Cotter$^{2}$,
T.\,J. Pearson$^{3}$, A.\,C.\,S. Readhead$^{3}$, 
\newauthor R.\,C.  Bolton$^{4}$, C.\, Chandler$^{5}$, G.\, Pooley$^{4}$, J.\, M.\, Riley$^{4}$ and E.\, M.\, Waldram$^{4}$\\
$^1$ Department of Physics \& Astronomy, University of Sheffield, Sheffield S3 7RH\\
$^2$ Astrophysics, Denys Wilkinson Building, Keble Road, Oxford OX1 3RH\\
$^3$ California Institute of Technology, Pasadena, CA 91125, USA\\
$^4$ Astrophysics, Cavendish Laboratory, J J Thomson Avenue, Cambridge CB3 0HE\\
$^5$ NRAO VLA, Array Operations Center, PO Box O, 1003 Lopezville Road, Socorro NM 87801-0387, USA}

\begin{document}

\maketitle

\date{Draft version 31 Mar 2005}

\pagerange{\pageref{firstpage}--\pageref{lastpage}}

\pubyear{2005}

\label{firstpage}

\begin{abstract}

9C J1503+4528 is a very young CSS radio galaxy, with an age of order
$10^{4}$ years.  This source is therefore an ideal laboratory for the study of
the intrinsic host galaxy/IGM properties, interactions between the
radio source and surrounding ISM, links between star formation and AGN
activity and the radio source triggering
mechanism. Here we present the results of a spectroscopic analysis of
this source, considering each of these aspects of radio source
physics.

We find that shock ionization by the young radio source is important
in the central regions of the galaxy on scales similar to that of the
radio source itself, whilst evidence for an AGN ionization cone is
observed at greater distances.  Line and continuum features require
the presence of a young stellar population, the best-fit model for
which implies an age of $5 \times 10^6$ years, significantly older
than the radio source. 

Most interestingly, the relative sizes of radio source and extended
emission line region suggest that both AGN and radio source are
triggered at approximately the same time.

If both the triggering of the radio source activity and the formation
of the young stellar population had the same underlying cause, this
source provides a sequence for the events surrounding the triggering
process. We propose that the AGN activity in 9C J1503+4528 was caused
by a relatively minor interaction, and that a super-massive
black hole powering the radio jets must have been in place before the
AGN was triggered.

\end{abstract}


\section{Introduction}

\footnotetext{E-mail: k.inskip@shef.ac.uk}

The triggering of powerful radio sources is still poorly
understood. It is generally thought that some mechanism must channel
significant quantities of gas into the central regions of the host to
fuel the AGN, wherein a rapidly spinning black hole may power the
radio jets (e.g Blandford and Znajek, 1977).

Two scenarios have been proposed. The first is the infall of gas which
has cooled in the halo
of a galaxy, or a cooling flow in a cluster (Bremer, Fabian \&
Crawford 1997); this proposition has received much interest recently
because it may have a thermostat feedback effect to prevent runaway
cooling in clusters (Br\"uggen and Kaiser 2002, Fabian et
al. 2003). However it is most likely that this mechanism in fact
causes a renewal of activity in an existing radio source (Nipoti and
Binney 2005), rather than being related to the original triggering.

The second is associated with the long-standing idea that a merger
between two galaxies, at least one of which is gas rich, is a
precursor to AGN activity (e.g. Heckman et al 1986). Wilson and
Colbert (1995) suggested that the coalescence of black holes following
a major merger creates the necessary rapidly-spinning black hole,
which suggests links between radio sources and other merger systems
such as ULIRGs, and implies that evidence for young stellar
populations and disturbed morphologies should be present at some level
in observations of young radio sources.

The merger scenario is supported by an increasing amount of anecdotal
evidence. For instance, there are several cases known where the host
galaxy of a 
powerful radio source appears to be undergoing a major merger (e.g., Johnston
et al. 2005). Similarly, there is increasing evidence that
powerful high-redshift radio galaxies lie in forming clusters (see
e.g. Venemans et al. 2005); such environments would be rich in major
mergers.  This evidence in turn takes us back to the first scenario
above; massive cluster galaxies may host a spinning black hole as a
relic of their early history, which is generally quiescent but which
becomes active when fuelled by cooling gas. It is clearly evident that
radio sources play a significant part in 
hierarchical galaxy and structure formation (e.g. Croton et al 2006), so the precise
ingredients for radio source formation are a cause of renewed
interest.

In any case, our understanding of this early phase of radio galaxy
evolution is hampered by the difficulty of compiling uniform samples
of radio sources in the first few millennia of their growth, and
indeed the impossibility of pinpointing galaxies immediately prior to
this stage.  However, it is is now widely accepted that Compact Steep
Spectrum (CSS) and some Gigahertz Peaked Spectrum (GPS) radio sources
represent the earliest stages of radio source evolution, being the
progenitors of 
the larger FRII source population (e.g. Fanti 1995, Readhead 1996,
Snellen et al 2000, Murgia 2003).
The study of such sources in the stages immediately after
triggering may be a valuable way of elucidating the events that have
occurred beforehand.

In an effort to secure a uniform sample of young sources, we have
turned to the 9C survey (Waldram et al 2003). This was carried out at
15 GHz with the Ryle Telescope to identify foreground sources for the
Very Small Array (VSA) microwave background experiment.  Over the past
several years, extensive followup of the radio properties of this
population has been carried out (Bolton et al 2003, 2004, 2006; Bolton et al
submitted), both to understand the effects on microwave
background experiments better and to investigate the sources themselves; the
high survey frequency has proved a good way of selecting significant numbers of CSS
and GPS sources, as well as sources peaking in the $\sim 10$ GHz
range. We are now undertaking optical followup of the CSS and GPS
sources, and in this paper we present a detailed spectroscopic study
of one of these compact sources, 9C J1503+4528, a classical double
FRII radio galaxy at a redshift of $z \sim 0.5$ with an angular size of
$0.5^{\prime\prime}$.

Details of the observations are presented in section 2, analysis of
the emission line strengths and ionization state in section 3, gas
kinematics in section 4, emission morphology in section 5 and a
discussion of the implications of our results in section 6.  We
give a summary of our conclusions in section 7.

Values for the cosmological parameters of $\Omega_0 = 0.27$,
$\Omega_{\Lambda} = 0.73$ and $H_0 = 65$km s$^{-1} \rm{Mpc}^{-1}$ are assumed
throughout.

\section{The Observations}

\begin{figure*}
\vspace{3.5 in}
\begin{center}
\includegraphics{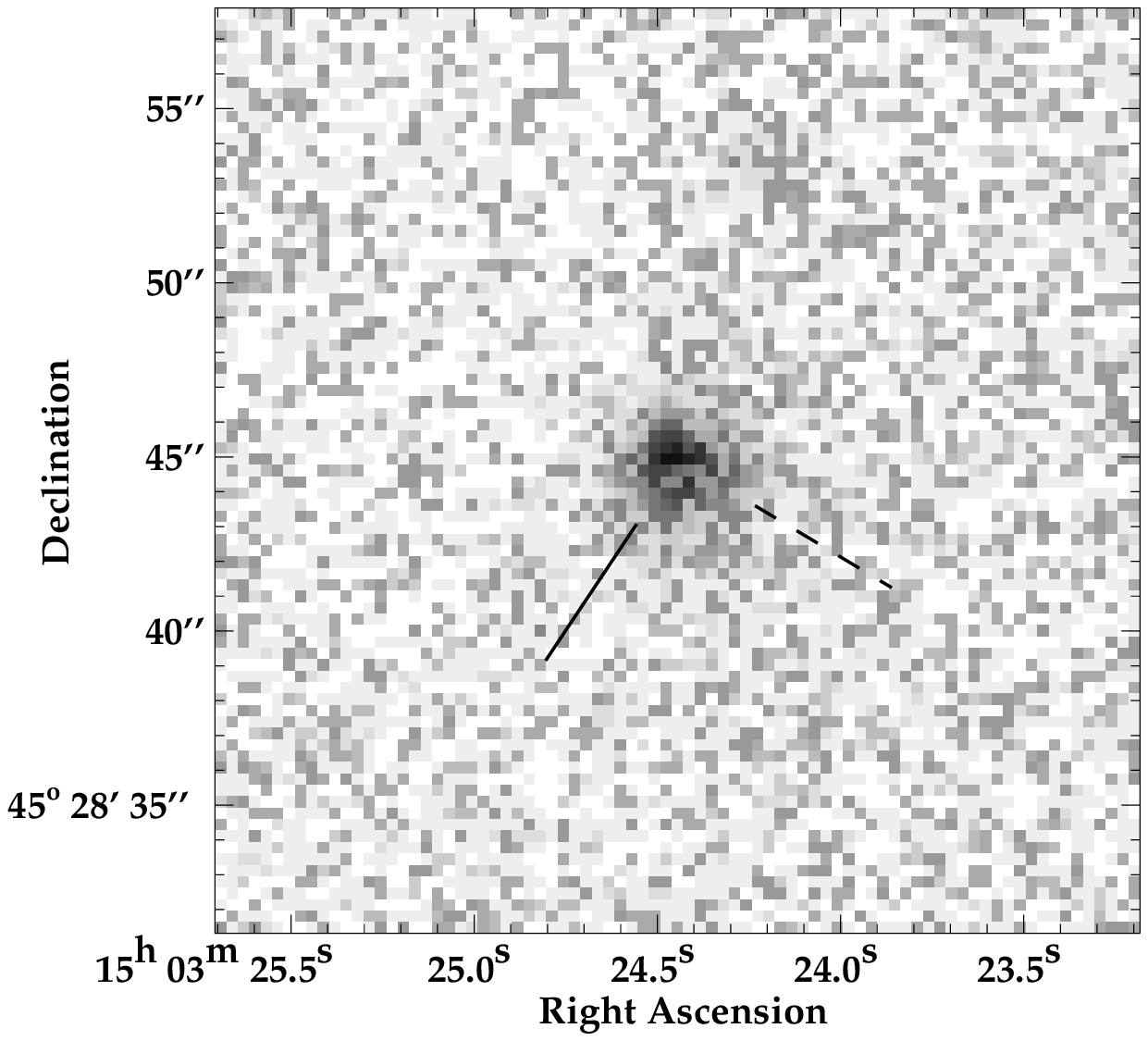}
\includegraphics{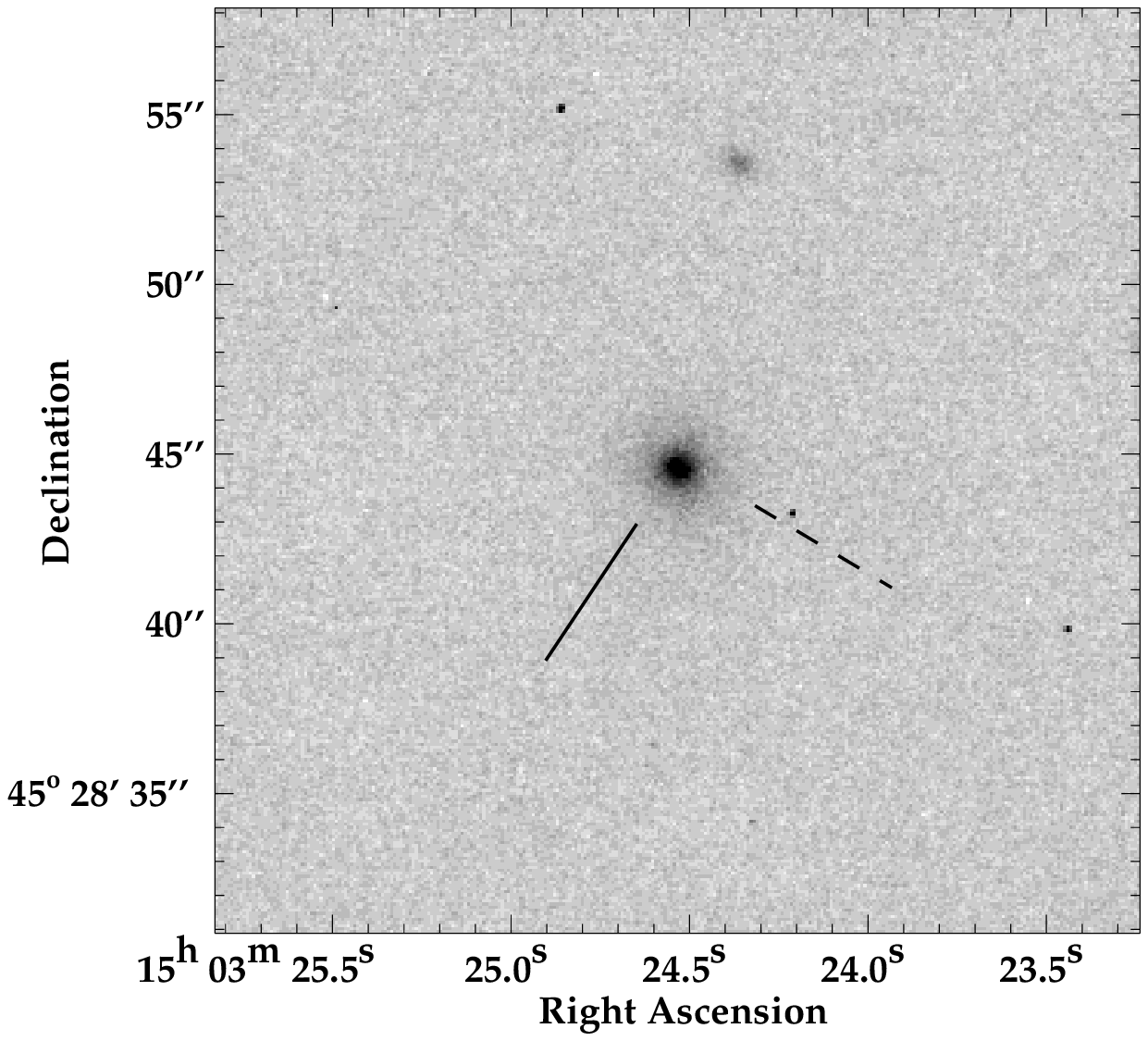}
\includegraphics{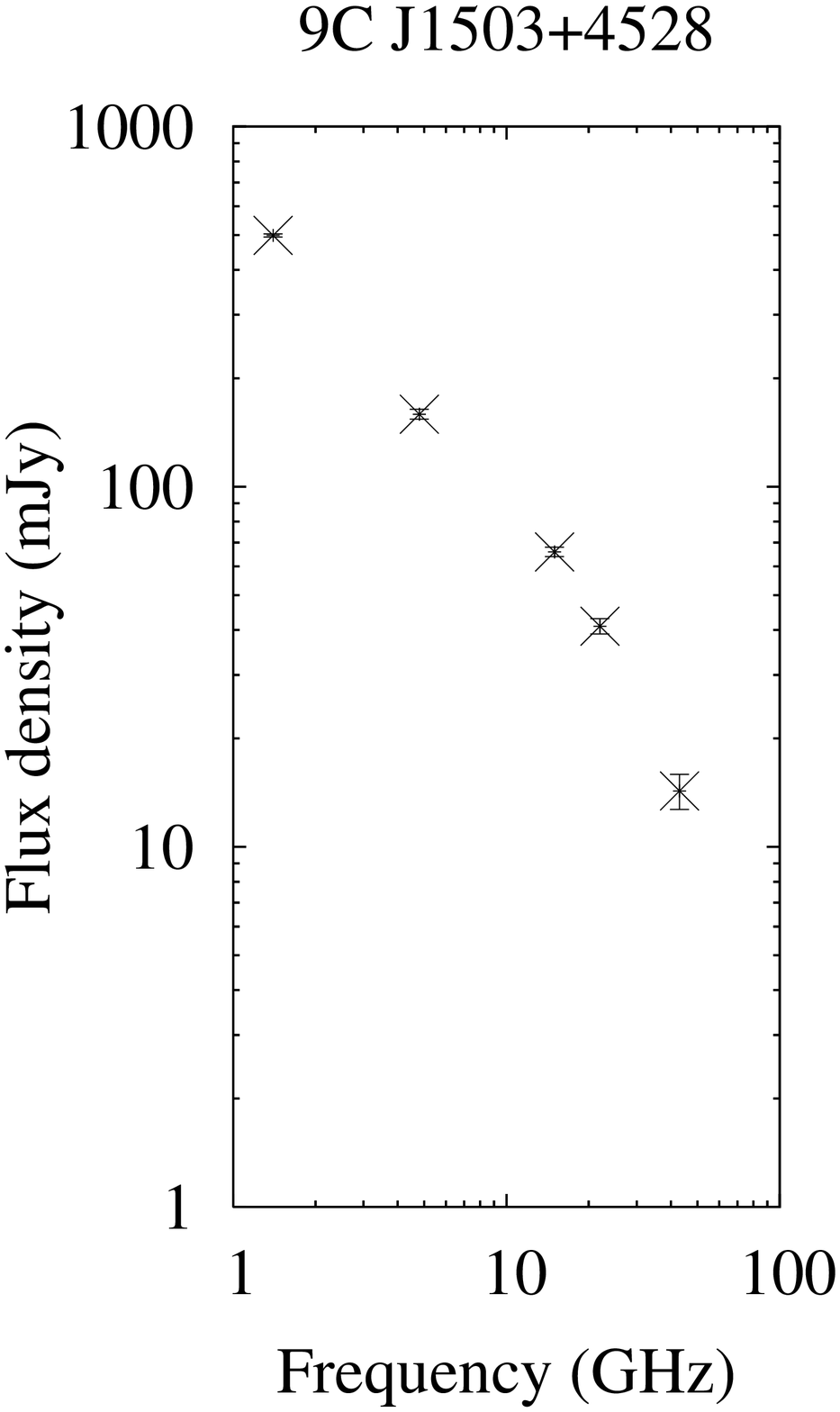}
\includegraphics{fig1d.eps}
\end{center}
\caption{The compact radio source 9C J1503+4528.  Left: UFTI $K-$band (top) and WFC $r-$band (bottom) images
  of the host galaxy, an $R \approx 19$ elliptical. The
  two spectroscopic slit PAs, parallel and perpendicular to the radio source axis
  are indicated on each diagram by the solid and dashed lines respectively. Centre: VLA 22GHz
  radio contours.     Right: VLA and OVRO radio spectrum for
  this source, from Bolton et al (2004). 
\label{PAfig}}
\end{figure*}

The observed source, 9C1503+4528 (see Fig.~\ref{PAfig}), was selected
as our first target for detailed study for several reasons.

First, the source has a rest-frame 1.4 GHz luminosity of 1.33 $\times
10^{25}$ W Hz sr$^{-1}$, well above the FRI/FRII break. Second, the
source is resolved, and clearly shows a classical double structure in the radio map presented
in Bolton et al. (2004). The angular size of the source is
$0.5^{\prime\prime}$, corresponding to a projected linear size of
$\approx 3$ kpc. It is currently accepted that GPS and CSS sources
are at a young stage in the evolution of the radio source (see e.g.
Snellen et al 2000). The picture in which sources are young is strongly
supported by observations of the hotspot advance speed in many of the
most compact sources.  These
observations find hotspot advance speeds in the range 0.1--0.4$c$
(e.g.  Polatidis and Conway 2003; Taylor et al 2000; Owsianik, Conway \& Polatidis
1998), which 
in the case of 9C J1503+4528 implies that the source has an age of order $\sim
10^4$ years. 
 This places the source very close to its time of
triggering, since the typical lifetime of an FRII source is $\sim
10^7$ to $10^8$ years (e.g. Kaiser, Dennett-Thorpe and Alexander
1997).  
 Finally, the source redshift ($z \sim 0.5$) makes many useful diagnostic
lines and the 4000-\AA\ break accessible to moderate-resolution
spectroscopy in a wavelength range region relatively clear of night
sky emission.  With these facts in mind, the young CSS radio source 9C
J1503+4528 is therefore a particularly appropriate object for the
study of
the intrinsic host galaxy/IGM properties, interactions between the
radio source and surrounding ISM, links between star formation and AGN
activity and the underlying radio source triggering
mechanism.

\subsection{VLA imaging}

The 22-GHz data obtained by Bolton at el. (2004) were re-analysed to
obtain the map shown in Fig.~\ref{PAfig}. 
Both the central component and the SE compact component are unresolved.
The most obvious interpretation of the radio morphology is that the
bright central component is the core, with the SE compact component
being a hotspot, and the SE and NW extensions being lobes. Given the
overall steep spectrum, it is implausible that the source is being viewed
at an angle close to the line of sight; in such a case, a strongly
boosted flat-spectrum jet component would be expected to dominate in this
high-frequency map. We thus conclude that the source has a double-lobed
structure with an overall extent of $\sim 3$kpc.

\subsection{Optical and infrared imaging}

An image using the Isaac Newton telescope Wide Field Camera $r$-band
filter was taken on the night of 2005 April 11. This filter has the
advantage of not including the strong {\sc [Oii]} and {\sc [Oiii]}
emission lines at the source redshift.  Five 600-s images were taken,
offset at the centre and corners of a 20$^{\prime\prime}$ pattern, and
stacked. The resulting image is shown in Fig.~\ref{PAfig}.

\begin{figure}
\vspace{6.6 in}
\begin{center}
\includegraphics{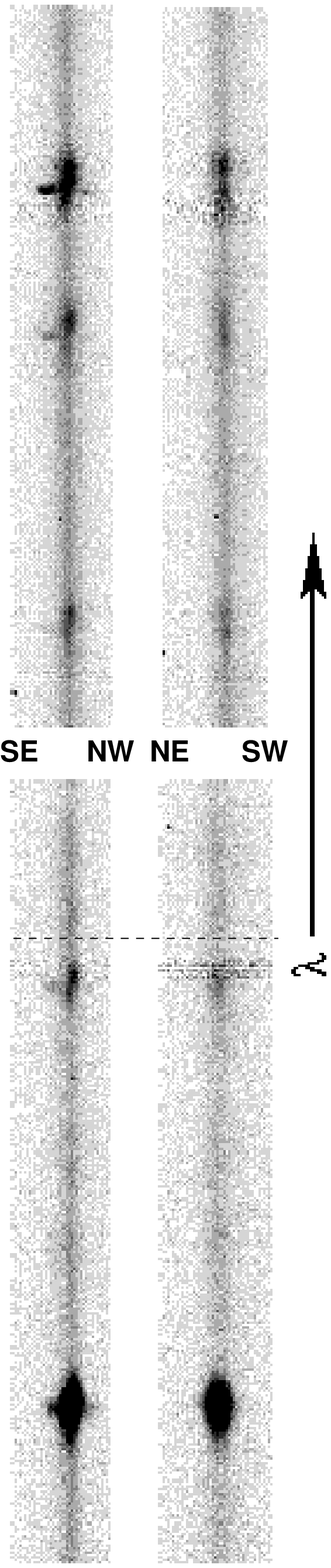}
\end{center}
\caption{Sections of particular interest (rest frame 3730-4070\AA\ and 5190-5490\AA) from the 2-d optical spectra
  of 9C1503+4528, with the slit aligned 
  parallel (left) and perpendicular (right) to the radio source
  axis.  Wavelength increases from the bottom to the top of the
  diagram, as indicated by the direction of the arrow, and the
  position of the 4000\AA\ break is denoted by the dashed line.  The
  extracted regions are approximately $10^{\prime\prime}$ wide.  The five
  prominent emission lines are (bottom to top): \oo3727\AA, [Ne\textsc{iii}]3869\AA, H$\beta$, \o34959\AA\
  and \o35007\AA.
\label{2dspec}}
\end{figure}

A $K$-band image was also obtained, on the night of 2006 March 5 using
the UFTI instrument at UKIRT. Nine 60-s jittered observations were
taken and reduced 
using the standard {\sc orac-dr} pipeline; the resulting image is also
displayed in Fig.~\ref{PAfig}. The
seeing, measured from an observation of UKIRT standard star P177-D
taken immediately after the target observation, was 0.6 arcseconds.

\subsection{Optical Spectroscopy}

The spectroscopic observations were carried out on 2004 July 18, using
the Keck LRIS spectrograph (Oke et al 1995).  Seeing was $\sim
0.8^{\prime\prime}$, and a $1^{\prime\prime}$ slit was used.  The
600/7500 grating was used, providing a wavelength range of $\sim
5200-7800$\AA, a spatial scale of 0.215 arcsec/pixel, a spectral scale
of $1.26$\AA/pixel, and a spectral resolution of $\sim 4.4$\AA.  For
an object lying at a redshift of $z \sim 0.5$, the corresponding
rest-frame spectral range is approximately $3500-5200$\AA, thereby
including emission lines from [O\textsc{ii}]3727\AA\ up to
[O\textsc{iii}]5007\AA.  Slit PAs of $+45^{\circ}$ and $-45^{\circ}$
were chosen, aligning the slit either parallel or perpendicular to the
radio source axis.  Long-slit spectra were obtained with total
integration times of 1800s at each slit PA. The integrations were split into 900-s exposures
to aid the removal of cosmic rays.

Standard packages within the NOAO \textsc{iraf} reduction software
were used to reduce the raw data. Corrections were made for overscan
bias subtraction, and the data were then flat-fielded using internal
lamps.  The flat fields were taken immediately after each observation
of the target, to minimize the effects of fringe drift in the CCD.
The two-dimensional frames were wavelength calibrated and corrected
for distortions in the spatial direction using wavelength-calibration
arc exposures. Taking care not to subtract any of the extended line
emission, the sky background was removed.  Observations of the
smooth-spectrum standard star BD +28 4211 were used to provide accurate
flux calibration of the final spectra and to remove the telluric A- and
B-band absorption features. The central regions of the parallel and
perpendicular spectra
effectively probe the same region of the host galaxy. These two
spectra were combined in order to produce a higher signal-to-noise
spectrum, for use in analysing the
emission from the central regions of the radio galaxy.

\begin{figure}
\vspace{2.1 in}
\begin{center}
\includegraphics{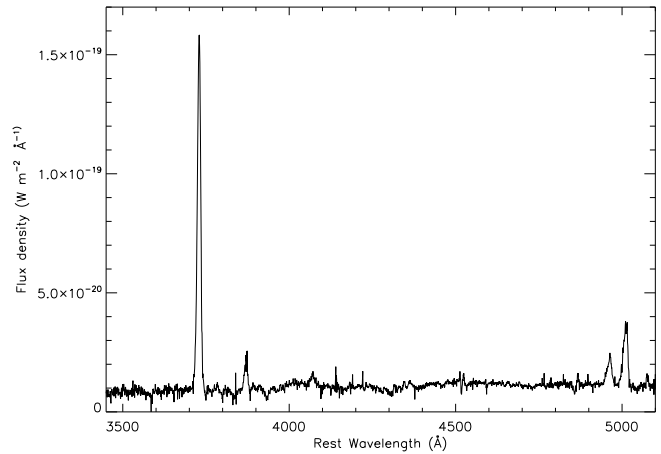}
\end{center}
\caption{Spectrum of 9C J1503+4528 obtained from the central 2\arcsec\
  of the combined parallel and perpendicular data.  This spectrum
  provides the highest signal-to-noise ratio for our data, and
  provides a good illustration of the relative emission line strengths
  and the fairly blue colour of the continuum.
\label{nicespec}}
\end{figure}

The resulting parallel and perpendicular 2-d spectra are displayed in
Fig.~\ref{2dspec}.  In the parallel spectrum, the presence of narrow
emission, extended over a distance of $\sim 5^{\prime\prime}$
(approximately 10 times larger than the radio source itself) can
clearly be seen, in addition to a broader component lying close to
the continuum centroid.  The \oo\ emission is particularly strong
compared to \o3, and the continuum emission short of the 4000\AA\
break is also substantially more luminous than would be expected for
an old stellar population alone.  The perpendicular spectrum displays similar properties, with
the exception that extended narrow emission is not observed.  A 1-d
spectrum extracted from the central 2\arcsec\ of the combined
perpendicular and parallel spectra is displayed in
Fig.~\ref{nicespec}.

Determining the precise systemic redshift of 9C J1503+4528 was
complicated by the inherent structure in the emission lines: the
redshift implied by the peak of the line emission on the continuum
centroid differs substantially from that of the narrow extended
component.  However, the various absorption features in the spectra
(Ca H \& K, high-order Balmer lines) all suggest a redshift identical
to that determined from the extended narrow component, giving a value
of $z = 0.521$.

\section{Emission line fluxes and the ionization state of the gas}

One-dimensional spectra were extracted from the parallel,
perpendicular and combined 2-d frames. The extractions were carried out
over $2^{\prime\prime}$ ($\sim 12$kpc) for the parallel,
perpendicular and combined frames, and
additionally over $5^{\prime\prime}$ ($\sim 31$kpc) for the parallel
and perpendicular frames. The lines were then fitted by gaussians; the
resulting fluxes and equivalent widths are presented in Appendix A.

Spectra were also extracted in $0.5^{\prime\prime}$ steps across the
slit, in order to investigate the changing ionization state of the
emission line gas with position via ionization state diagnostic
diagrams (Baldwin, Phillips \& Terlevich 1981).   For an effective
analysis, model predictions should be well separated and results
should be consistent over several different diagrams/line
combinations.  Line pairs should ideally be close in wavelength and of
the same species, to limit the effects of reddening and variations in
gas composition. The most appropriate
lines available from our optical spectra are \oo3727\AA, \o34363\AA,
H$\beta$ and \o35007\AA.  

\begin{figure}
\vspace{2.3 in}
%\vspace{7.45 in}
\begin{center}
\includegraphics{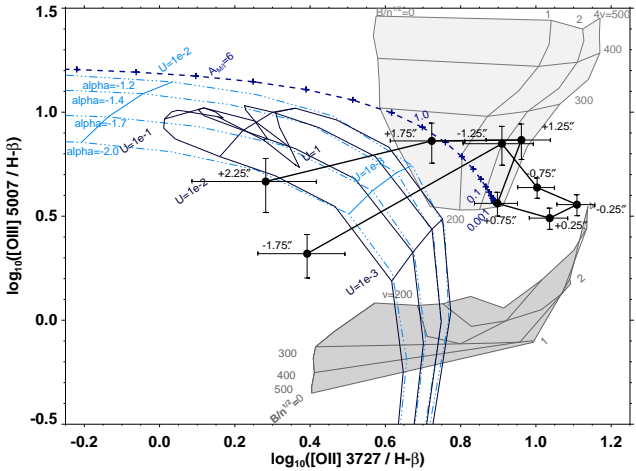}
%\special{psfile='fig4a.eps' hoffset=260 voffset=+335 angle=90 
%hscale=37 vscale=37}
%\special{psfile='fig4b.eps' hoffset=260 voffset=+150 angle=90 
%hscale=37 vscale=37}
%\special{psfile='fig4c.eps' hoffset=260 voffset=-35 angle=90 
%hscale=37 vscale=37}
\end{center}
\caption{Ionization mechanism diagnostic plot for 9C1503+4528, using
  the line ratios [O\textsc{ii}]3727/H$\beta$ and [O\textsc{iii}]5007/H$\beta$. Data
  points represent spectra extracted in 0.5$^{\prime\prime}$ steps,
  and are labelled in terms of their distance (in arcseconds) from the continuum
  centroid.  The model tracks  include: (1) shock 
  ionization (Dopita \& Sutherland 1996), with and
  without a precursor photoionization region (light \& dark shading
  respectively), (2) the matter-bounded photoionization track (dashed
  dark blue track, Binette,
  Wilson \& Storchi-Bergmann 1996), and (3) simple 
  AGN photoionization tracks (light blue tracks), including
  the effects of 
  dust (solid black tracks; Groves, Dopita \& Sutherland
  2004a,b). The photoionization models are for n=1000 cm$^{-3}$, solar
  metallicity and various values of the ionization index $\alpha$
  (from $1.2 < \alpha < 2.0$). A brief description of the effects of
  changes in metallicity and/or gas density is given in the main text; we
  refer the reader to Groves et al (2004a,b) for full details.
\label{ionize1}}
\end{figure}

Figure \ref{ionize1} displays the diagnostic diagram for the line
pairs \oo3727\AA/H$\beta$ and \o35007\AA/H$\beta$, 
extracted from the parallel spectra in 0.5$^{\prime\prime}$ steps,
together with various theoretical model tracks.  These include: (i)
the shock models of Dopita \& Sutherland (1996), which also
incorporate the effects of a precursor ionization field, (ii) the
mixed medium (matter bounded vs. ionization bounded) photoionization
tracks of Binette et al (1996), plus (iii) simple AGN photoionization
tracks (Groves et al 2004a, 2004b), also incorporating the effects of
dust.  The photoionization models displayed are for solar
metallicity and a density of $n=1000 \rm{cm^{-3}}$.  For reference, the
photoionization tracks display little variation for lower values of
$n$, while the $n=10000 \rm{cm^{-3}}$ track is shifted to smaller
values of log(\oo/H$\beta$) by $\sim 0.2$dex.  The $2Z_\odot$
metallicity tracks are generally similar to those of solar
metallicity, whereas lower $Z$ tracks are shifted to smaller values of
both log(\oo/H$\beta$) and log(\o3/H$\beta$).
 %Three different versions of this diagram are displayed,
%allowing the effects of spectral index, gas density and metallicity on
%the AGN photoionization tracks to be considered separately, and
%providing easier comparison between the other theoretical models and
%our observational data.

This diagram clearly shows the changing ionization state relative to
distance from the nucleus.  The spectra extracted from the central
regions of the host galaxy lie close to the predictions of the shock
ionization models (and also the far end of the mixed medium ionization
track).  At intermediate distances ($\sim 1^{\prime\prime}$, i.e. sampling emission from just
outside the radio source), the shock plus precursor photoionization
models are most suited.  Beyond 1-1.5$^{\prime\prime}$ (i.e. the
extended narrow emission region), the data are well described by
simple AGN photoionization.  High values of the ionization parameter,
$U$, are the most plausible, and a dusty medium is favoured.  For the
photoionized gas, the ionization state at larger radii suggests either
a lower spectral index ($\alpha \sim -2.0$ rather than $\alpha \sim
-1.2$), higher gas density or lower metallicity.  The mixed medium
models of Binette et al only provide a good explanation for a few of
the data points, and it seems unlikely that this system can be well
described using that particular model.    The observed line ratios
are not likely to be affected by internal reddening, as the Balmer
line ratios in our spectra are in good agreement with the predictions
of case B recombination.  Balmer line absorption could potentially
have some influence on the exact positioning of the data points in
this figure; while we believe we have correctly accounted for the
errors on our measurement of H-$\beta$, additional shifts of up to
0.2dex towards smaller line ratios cannot be completely ruled out.
However, this would not 
have any effect on our overall interpretation of the data.

The picture provided by these results is as follows.  The UV continuum
emission from the AGN ionizes the gas at all locations within the
ionization cone, decreasing in intensity at larger distances.  The
innermost regions of the galaxy, of order the same size as the
fledgling radio source, have an additional ionization contribution
from shocks.  The UV radiation field produced by these shocks adds
another ionization component to the emission from gas at intermediate
distances.

\begin{figure}
\vspace{2.3 in}
\begin{center}
\includegraphics{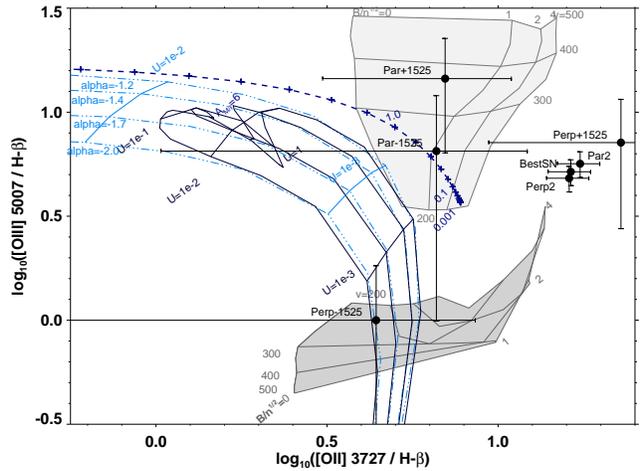}
\end{center}
\caption{Ionization mechanism diagnostic plots for 9C1503+4528, using
  the line ratios [O\textsc{ii}]3727/H$\beta$ and
  [O\textsc{iii}]5007/H$\beta$. Data 
  points represent spectra extracted from various regions of the
  spectra taken perpendicular and parallel to the radio axis, and
  include: (i) the central 2$^{\prime\prime}$ of the parallel,
  perpendicular \& combined (best signal-to-noise) spectra, (ii)
  1$^{\prime\prime}$ wide sections of the parallel and  perpendicular
  spectra lying $\pm 2^{\prime\prime}$ from the continuum centroid.  The
  model tracks are the same as those in Fig.~\ref{ionize1}a.  
\label{ionize2}}
\end{figure}

The relative ionization states of the parallel and perpendicular
spectra can be compared in Fig.~\ref{ionize2}.  This figure displays the
same tracks as in Fig.~\ref{ionize1}, for spectra extracted from
various regions of our 2-d spectra.  These include the central
$2^{\prime\prime}$ regions of the combined, parallel and perpendicular
spectra. The results for the central 1-2$^{\prime\prime}$ should not be
greatly different, as both spectra sample essentially the same
region. However, the impact of AGN and shock ionization will be
greatly lessened at larger distances, for material which lies off the
radio source axis and well outside the ionization cone of the AGN.
Also displayed are the observed emission line ratios for 
spectra extracted at distances from the continuum centroid of
$+1.5^{\prime\prime}$ to $+2.5^{\prime\prime}$ and $-1.5^{\prime\prime}$ to
$-2.5^{\prime\prime}$, for both the
parallel and perpendicular spectra.  The large errors on the data
points for the extended regions of the perpendicular spectra reflect
the very weak line emission present at these distances. In the case of
the  $-1.5^{\prime\prime}$ to
$-2.5^{\prime\prime}$ data point, shocks are not a plausible option,
and the data should lie outside the ionization cone of the AGN.
Instead, the position of this data point suggests that ionization by
young, hot stars may provide a better explanation (see e.g. Kewley,
Geller \& Jansen (2004) and Rola,
Terlevich \& Terlevich (1997) for appropriate locations on the
\oo/H$\beta$ vs. \o3/H$\beta$ diagnostic plot for
H\textsc{ii} regions; such emission line ratios would generally be
expected to lie towards the lower left quadrant of
Fig.~\ref{ionize2}).   A combination of ionization by young stars and
the AGN could also be relevant for the $-1.75^{\prime\prime}$ and
$+2.25^{\prime\prime}$ data points on Fig.~\ref{ionize1}.
 
\begin{figure}
\vspace{2.3 in}
\begin{center}
\includegraphics{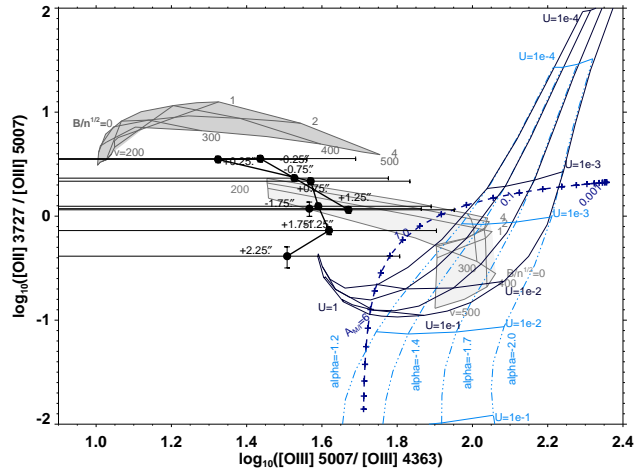}
\end{center}
\caption{Ionization mechanism diagnostic plots for 9C1503+4528, using
  the line ratios [O\textsc{ii}]3727/[O\textsc{iii}]5007 and
  [O\textsc{iii}]5007/[O\textsc{iii}]4363. Data 
  points represent spectra extracted in 0.5$^{\prime\prime}$ steps,
  and are labelled in terms of their distance from the continuum
  centroid.  The model tracks are the same as those in
  Fig.~\ref{ionize1}a, and include: (1) shock 
  ionization (Dopita \& Sutherland 1996), with and
  without a precursor photoionization region (light \& dark shading
  respectively), (2) the matter-bounded photoionization track (dashed
  dark blue track, Binette,
  Wilson \& Storchi-Bergmann 1996), and (3) Simple 
  AGN photoionization tracks (light blue, various styles), including
  the effects of 
  dust (solid black tracks, Groves, Dopita \& Sutherland
  2004a,b). 
\label{ionize3}}
\end{figure}

Figure \ref{ionize3} shows the diagnostic diagram for the
\o35007\AA/\o34363\AA\ and \oo3727\AA/\o35007\AA\ line ratios,
together with the observed data from our parallel spectra in spatial
steps of $0.5^{\prime\prime}$. Note that the mixed medium model track,
which bent towards the data points at low values of
$A_{M/I}$\footnote{The solid-angle ratio of matter-bounded to
  ionization-bounded material in the mixed-medium model of Binette et
  al 1996.} in
Fig.~\ref{ionize1}, tends towards higher values of
\o35007\AA/\o34363\AA\ on Fig.~\ref{ionize3} as opposed to the lower values of this line
ratio observed in the
data.   Although the error bars are large for the
\o35007\AA/\o34363\AA\ line ratio, we nevertheless have a certain
amount of consistency with the previous diagnostic diagrams: once
again the central regions of the galaxy favour the predictions of
shocks, while more distant locations lie closer to the dusty AGN
photoionization tracks.  The relatively low \o35007\AA/\o34363\AA\
ratio is indicative of high temperatures and/or densities in the
emitting gas, of the order of several $10^4$K in the low density limit
(e.g. Osterbrock 1989), fully consistent with the presence of shocks in
these regions.

\section{Gas Kinematics}

We have also investigated the changing gas kinematics, via an analysis
of the profiles of the \oo 3727\AA\ and \o3 (4959+5007\AA) lines.
Two-dimensional regions around each of these emission lines were
extracted, and from these a sequence of one-dimensional spectra were
extracted along the slit direction, stepped every
0.5$^{\prime\prime}$.

\begin{figure}
\vspace{3.7 in}
\begin{center}
\includegraphics{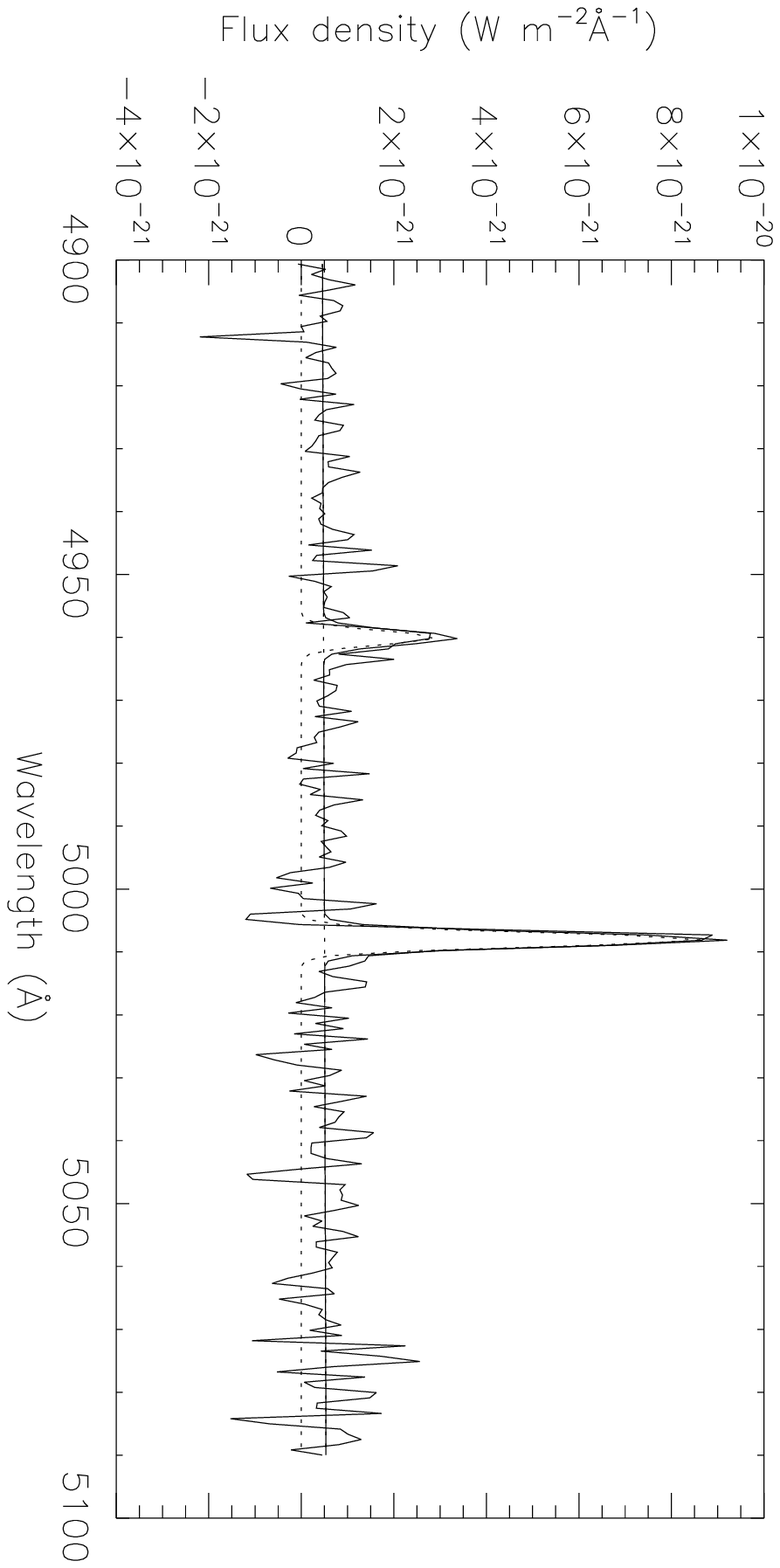}
\includegraphics{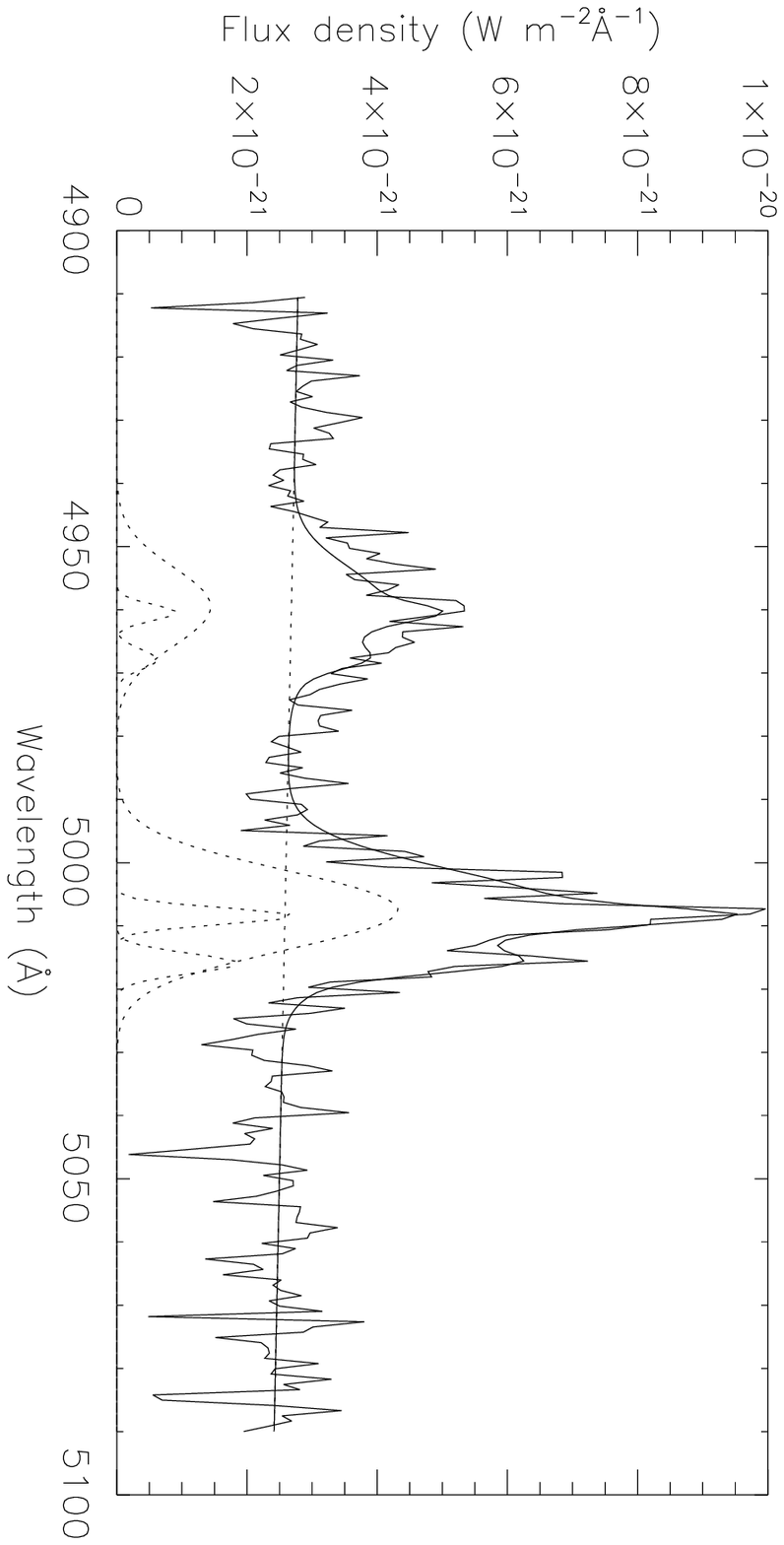}
\end{center}
\caption{Example line profile fits for the \o3(4959+5007)\AA\ emission
  lines.  The top figure is for the spectrum extracted at a distance
  of 2.25$^{\prime\prime}$ to the SE of the continuum centroid, and
  the bottom figure for a region lying at a distance of
  0.75$^{\prime\prime}$ in the same direction. The data were modelled
  with several components (dotted lines): a continuum level, and 1-3
  separate Gaussian components for each line.  For \o3, the relative
  strengths and positions of the  4959\AA\ and 5007\AA\ lines were
  fixed.  The combination of gaussians
with the lowest reduced $\chi^{2}$ determined the maximum
number of velocity components which could be fitted to
the data.
\label{kinfit}}
\end{figure}

After allowing for continuum subtraction, the data were fitted by up
to three successive gaussian components, which were only accepted if their FWHM
was larger than the instrumental resolution and the signal-to-noise
ratio was greater than approximately five (excepting the outermost
spectra, for which a fainter, single gaussian component could be
detected at a lower signal-to-noise.)  The combination of gaussians
with the lowest reduced $\chi^{2}$ were then taken to be the maximum
number of velocity components which could be fitted to
the data (we also fixed  the relative positions and
strengths of the 4959\AA\ and 5007\AA\ lines).
Example fits are displayed in Fig.~\ref{kinfit}.

\begin{figure}
\resizebox{\hsize}{!}
{\includegraphics[angle=-90]{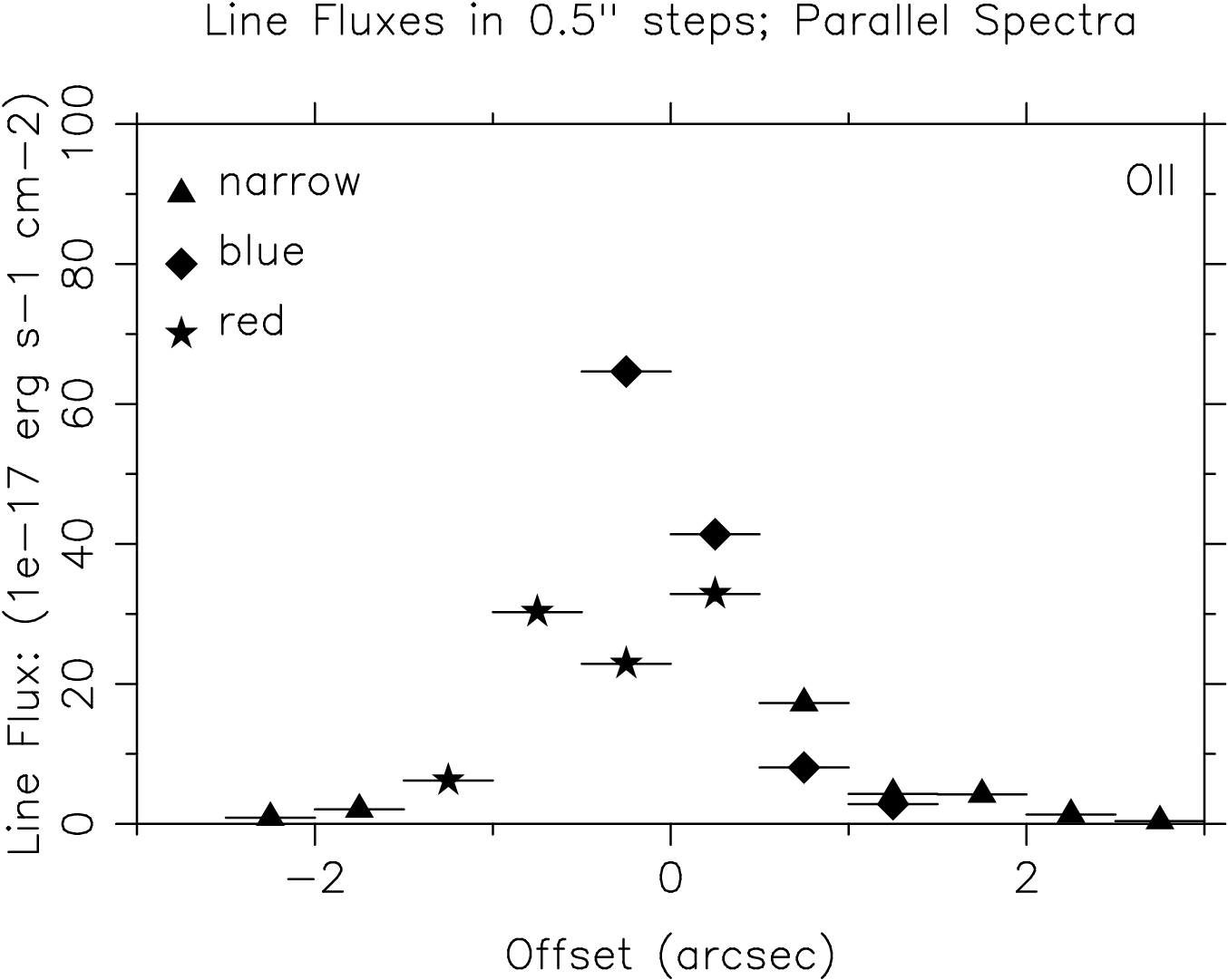}
\includegraphics[angle=-90]{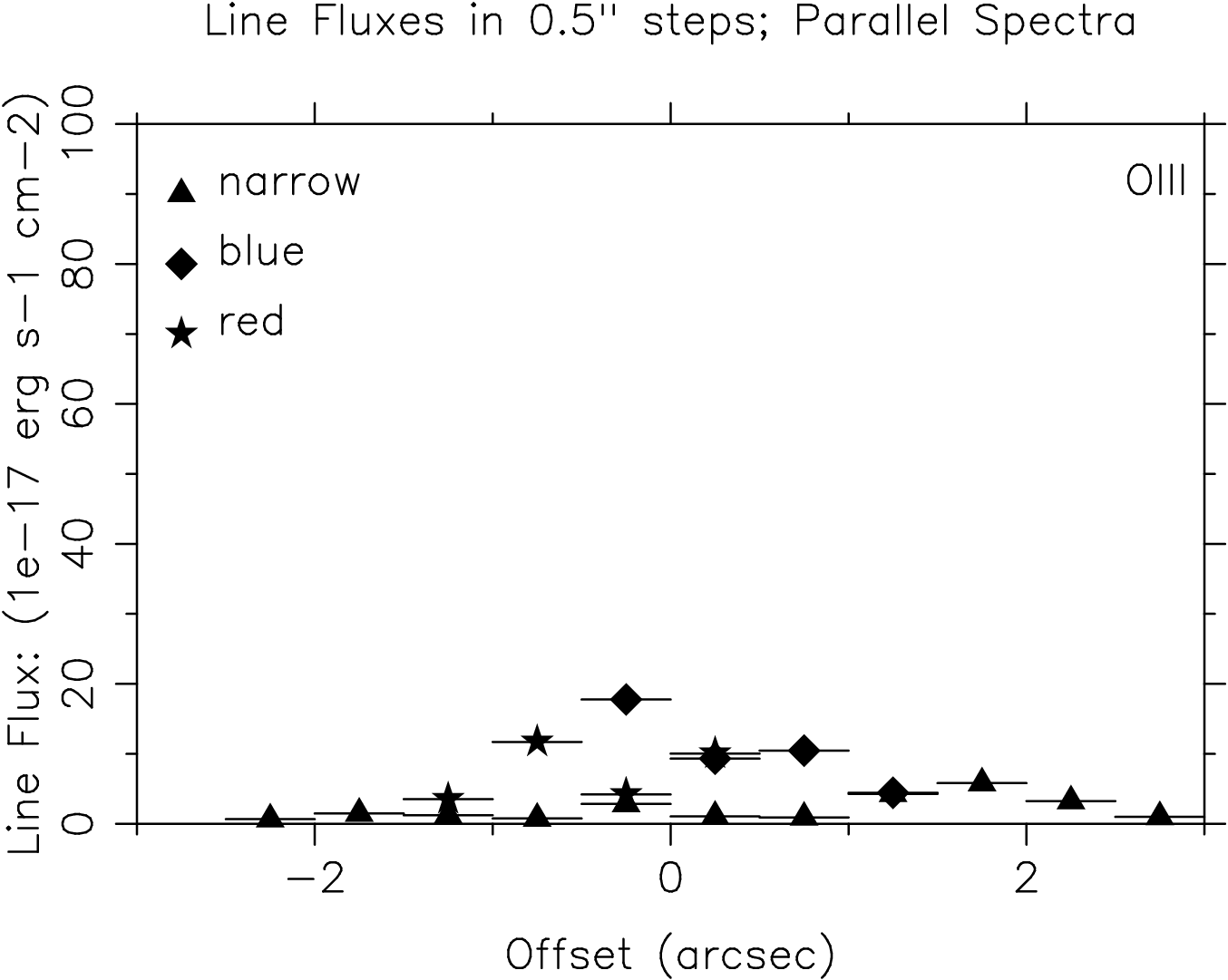}}\\
\resizebox{\hsize}{!}
{\includegraphics[angle=-90]{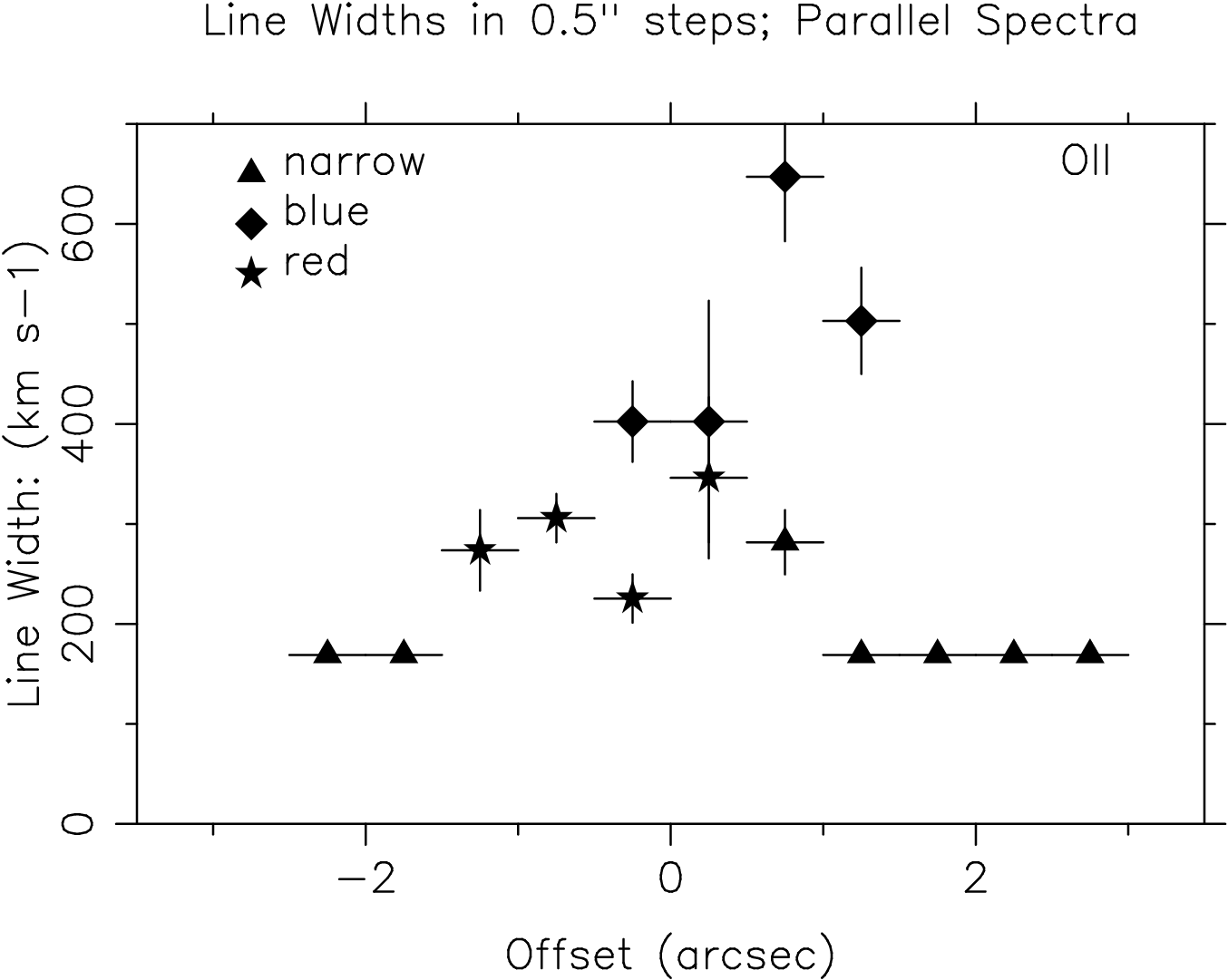}
\includegraphics[angle=-90]{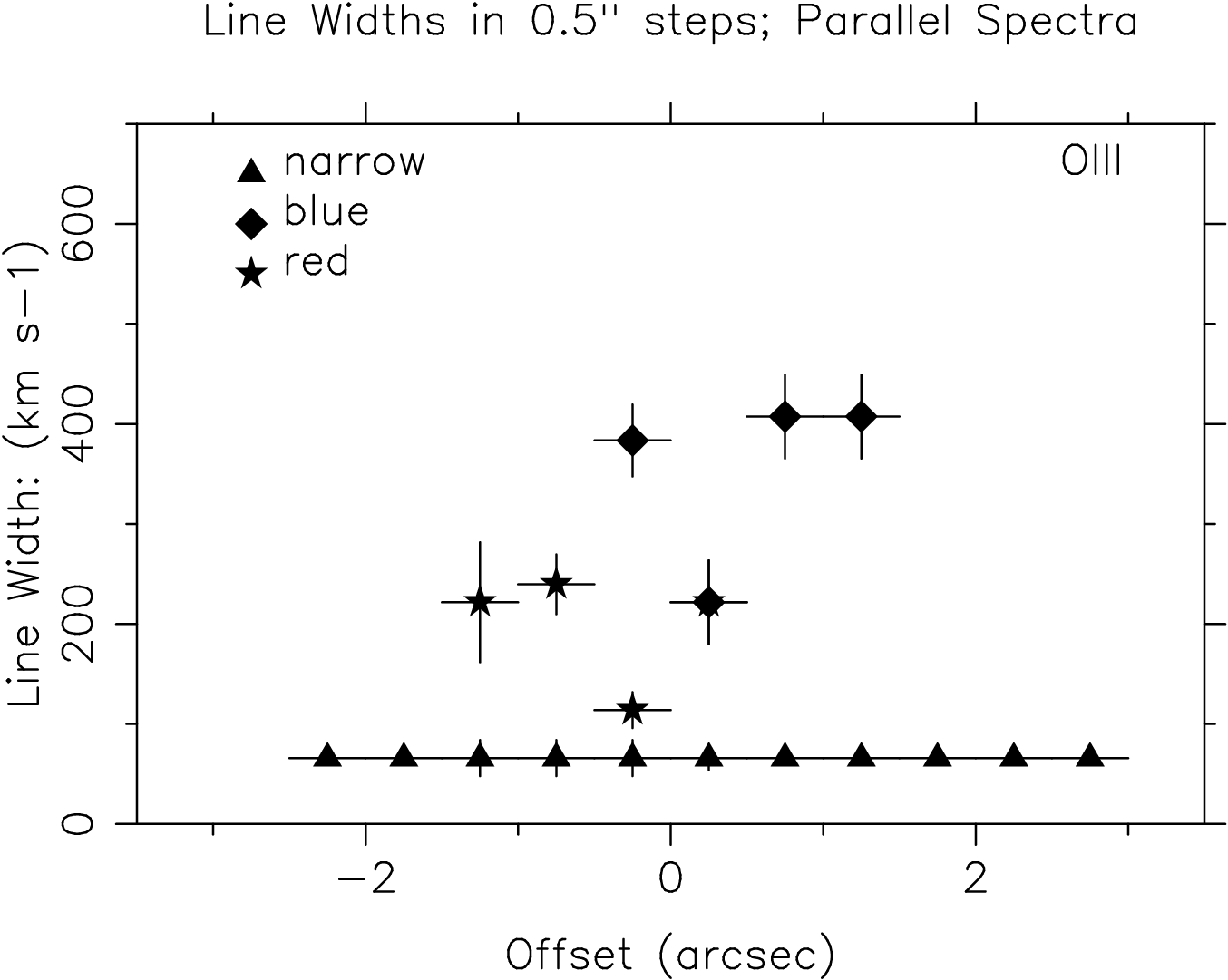}}\\
\resizebox{\hsize}{!}
{\includegraphics[angle=-90]{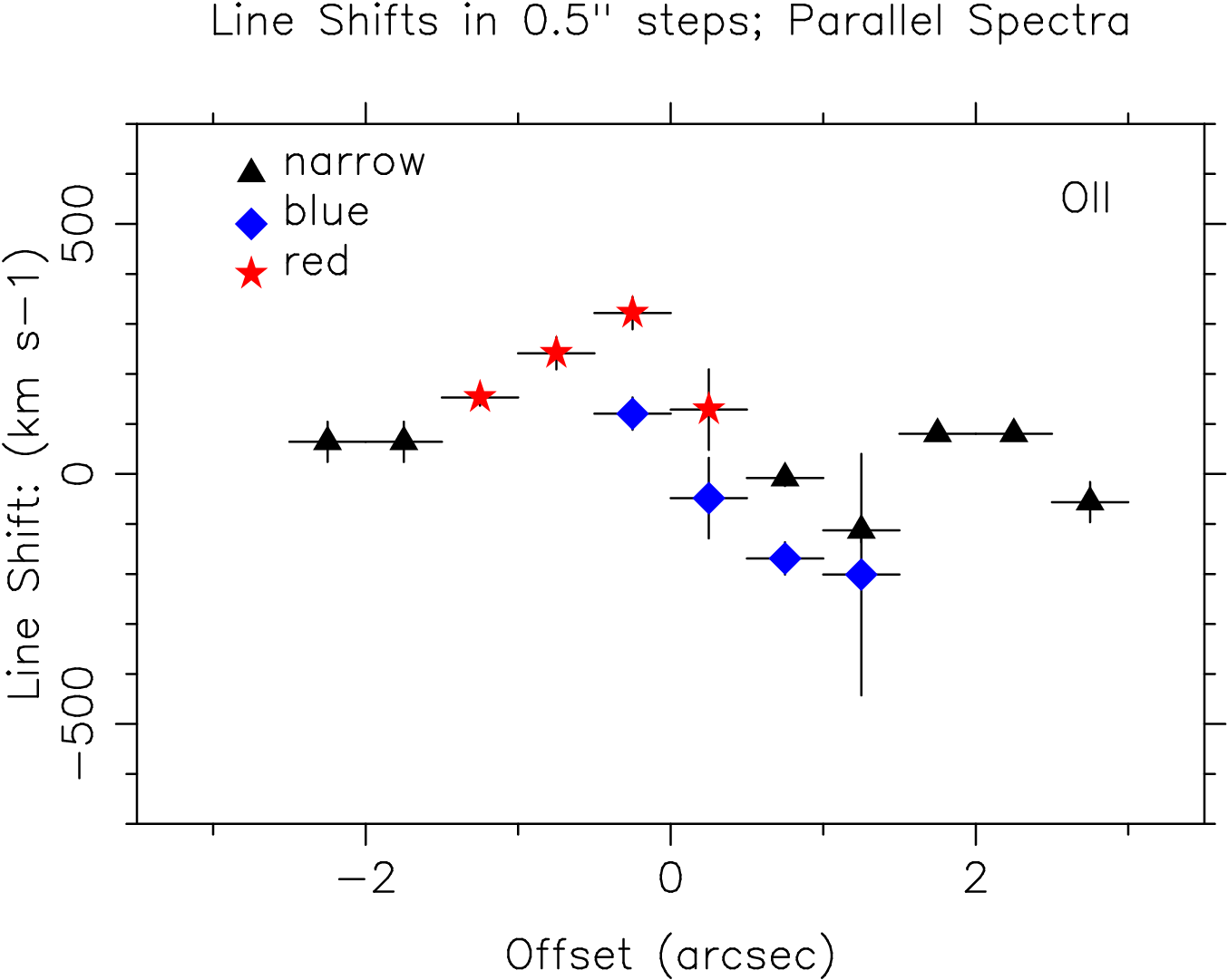}
\includegraphics[angle=-90]{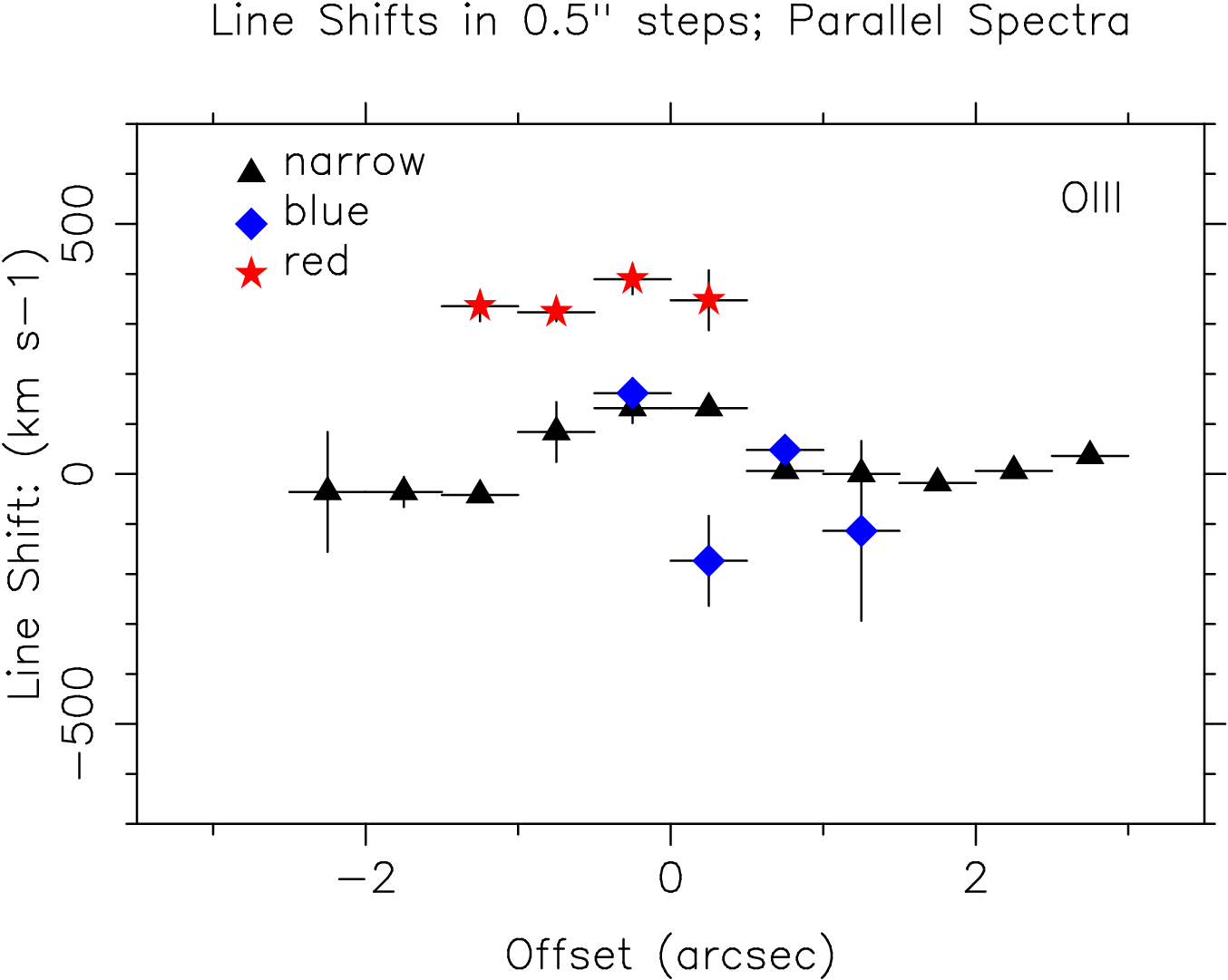}}
\caption{Gas kinematics for 9C J1503+4528.  The changing line fluxes
  (top), widths (centre) and offsets (bottom) with position are
  displayed for the \oo\ (left) and \o3\ (right) emission lines,
  extracted and fitted in 0.5\arcsec steps from the parallel
  spectrum. Up to three emission line components are modelled at each
  position.  Negative offsets are towards the NW, and positive offsets
  towards the SE.}
\label{K1}
\end{figure}

The integrated line flux, the velocity relative to that at the centre
of the galaxy, and the emission line FWHM were determined for each
gaussian fit. We calculated the FWHM by subtracting in quadrature the
instrumental FWHM, as determined from unblended sky lines. 
%This
%procedure assumes that the line emission illuminates the slit in a
%similar manner to the background sky emission, which is an adequate
%approximation for narrow slits and imperfect seeing conditions.
We then calculated errors for these three parameters, allowing for the
fact that a range of possible fits are equally acceptable at low S/N.
This approach allows us to search for high velocity components in the
emission line gas, or other structures incompatible with a fit to a
single velocity component.  Given the observed emission line profile
of this source, such a step was a necessity, and broader emission line
components are clearly present in the central parts of the extended
emission line region.

The results of our analysis of the kinematics of the parallel spectrum
are presented in Fig.~\ref{K1}. 

%, which displays the flux, width and
%velocity offset of up to three different line components at various
%positions across the spectrum, for both the \oo\ and \o3\ emission
%lines.

An unresolved emission component is present in the outer regions of
the extended emission line region (hereafter EELR), in both \oo\ and \o3, at wavelengths consistent with the
emitting material lying at rest relative to the host galaxy.  This
narrow component can be traced across the galaxy in \o3.  The \oo\
emission line is of substantially higher luminosity, particularly in
the central regions of the galaxy.  Because of this, the narrow
component of \oo\ can only be observed in the outer regions, and is
lost in the centre in favour of one or more broader components.
The relative strengths of the narrow components of
\oo\ and \o3\ are consistent with simple AGN photoionization, as
demonstrated by our diagnostic diagrams in the previous section.

The broader emission line components extend to approximately
1-1.5$^{\prime\prime}$ from the continuum centroid. Note that the \oo\
emission is substantially stronger than the \o3\ emission; such a low
ionization state in the emitting material is suggestive of shocks
associated with the radio source, in which case the emitting material
would display larger line widths than more settled material.  Usually a single broad component is
sufficient, except in the central regions where two broad components
are preferred.  The broad components on the SE side of the galaxy (see
the 2-d spectrum presented in Fig.~\ref{2dspec}) usually lie at
shorter wavelengths, with a velocity shift of a few hundred
km s$^{-1}$, whilst the emission from the opposite side of the galaxy
is shifted to longer wavelengths by a similar amount.  Of these two
distinct components, the blue-shifted emission line component is
generally more luminous, and has a larger line width (400-600
km s$^{-1}$, cf 200-400km s$^{-1}$ for the red-shifted emission).

\begin{figure}
\resizebox{\hsize}{!}
{\includegraphics[angle=-90]{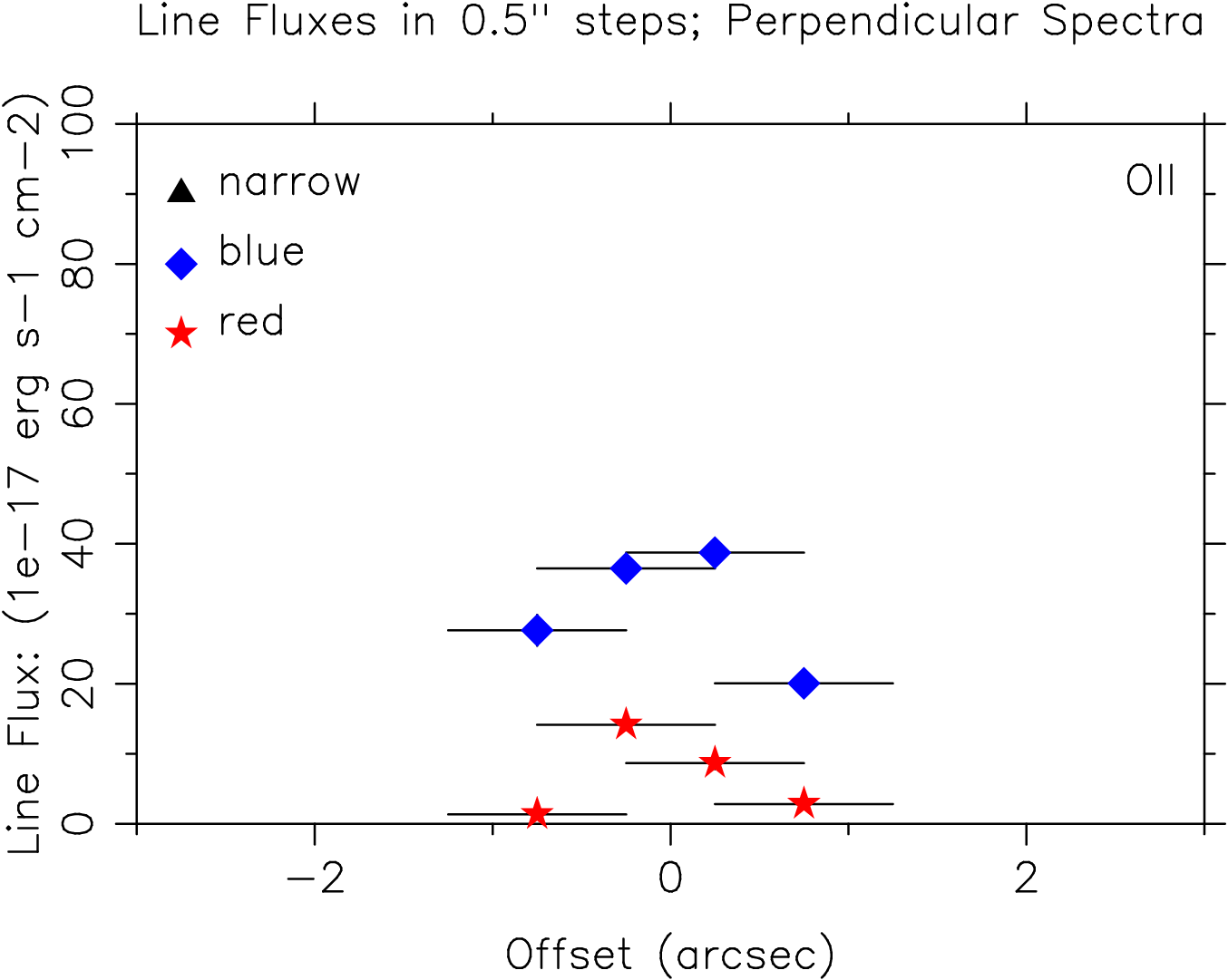}
\includegraphics[angle=-90]{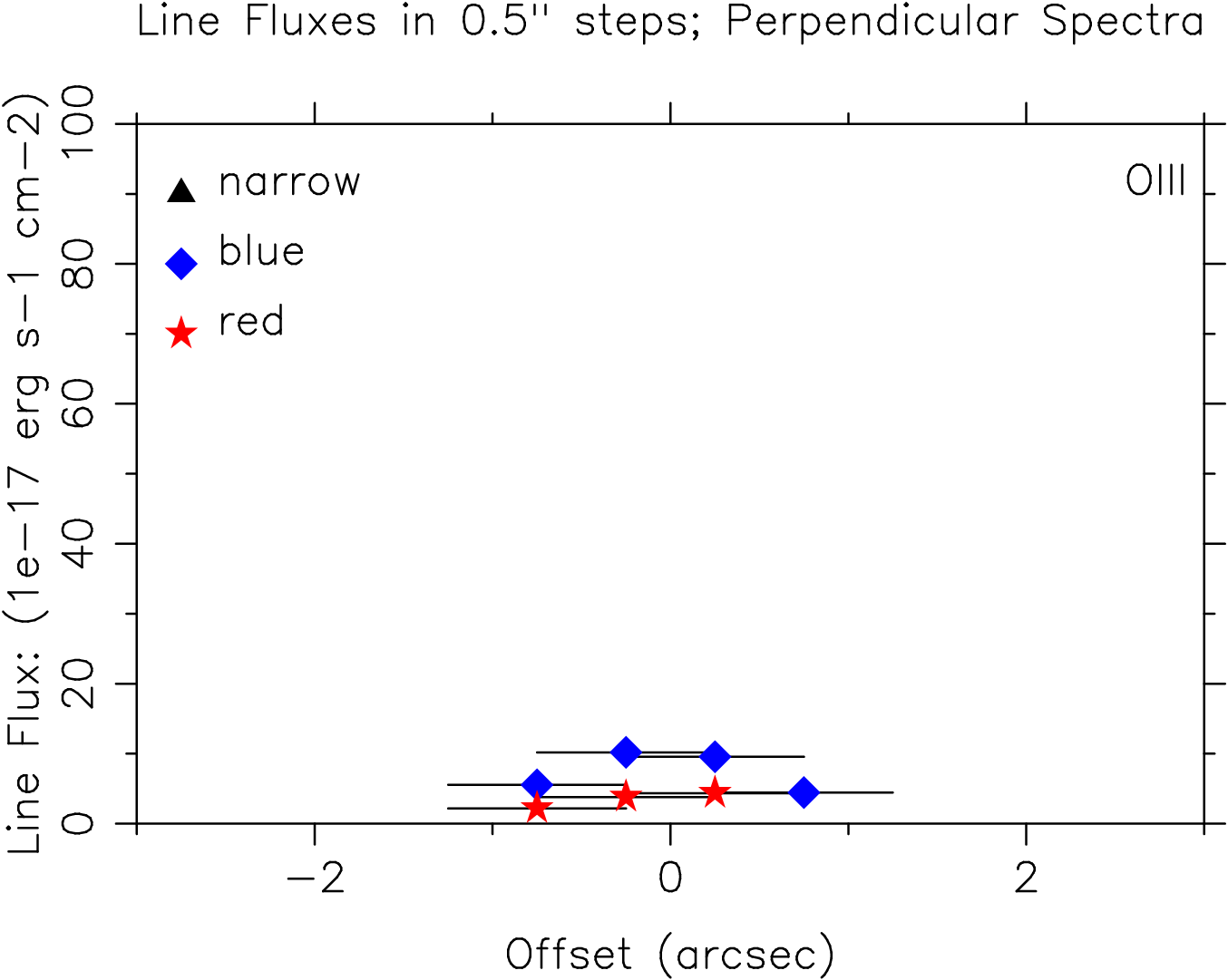}}\\
\resizebox{\hsize}{!}
{\includegraphics[angle=-90]{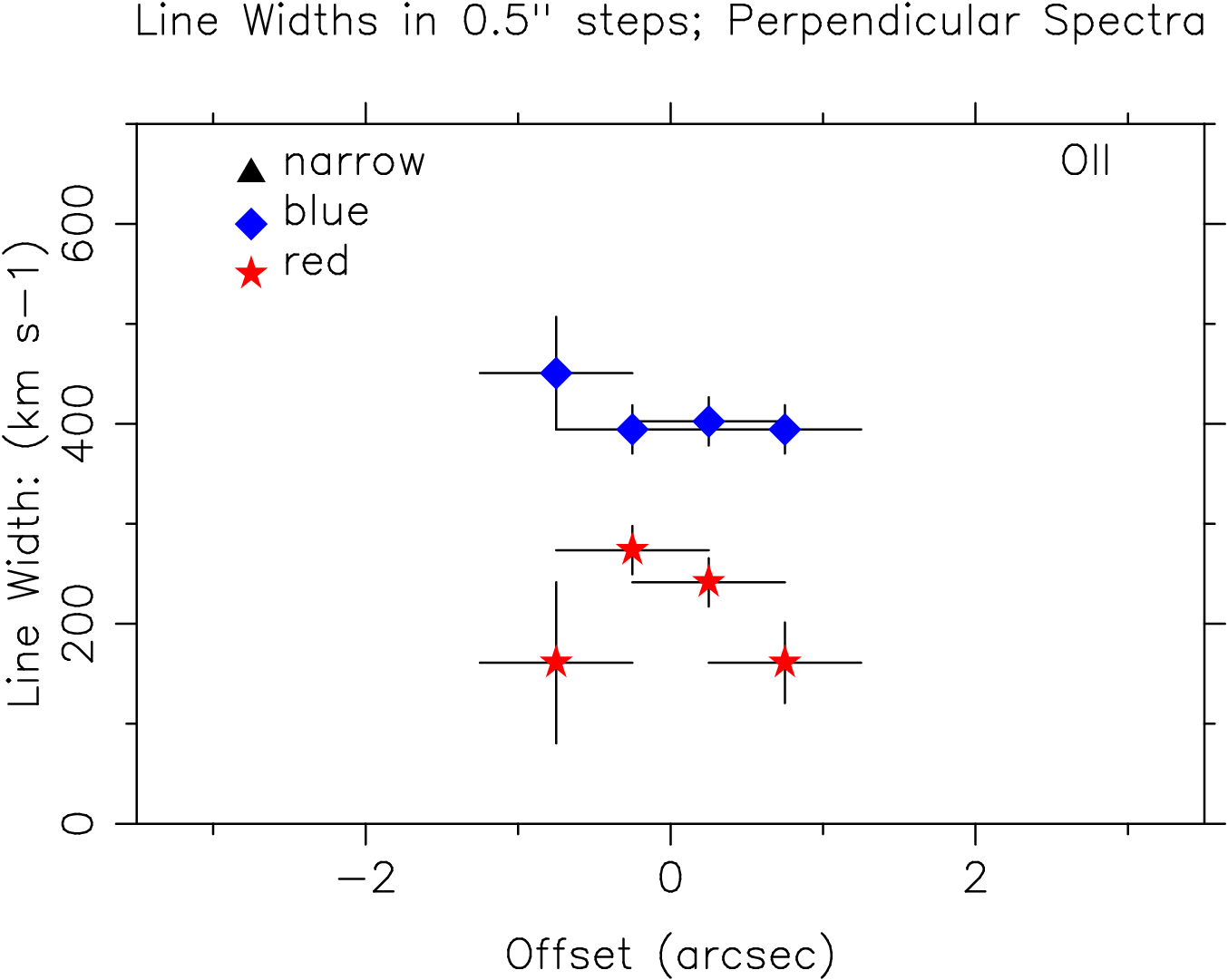}
\includegraphics[angle=-90]{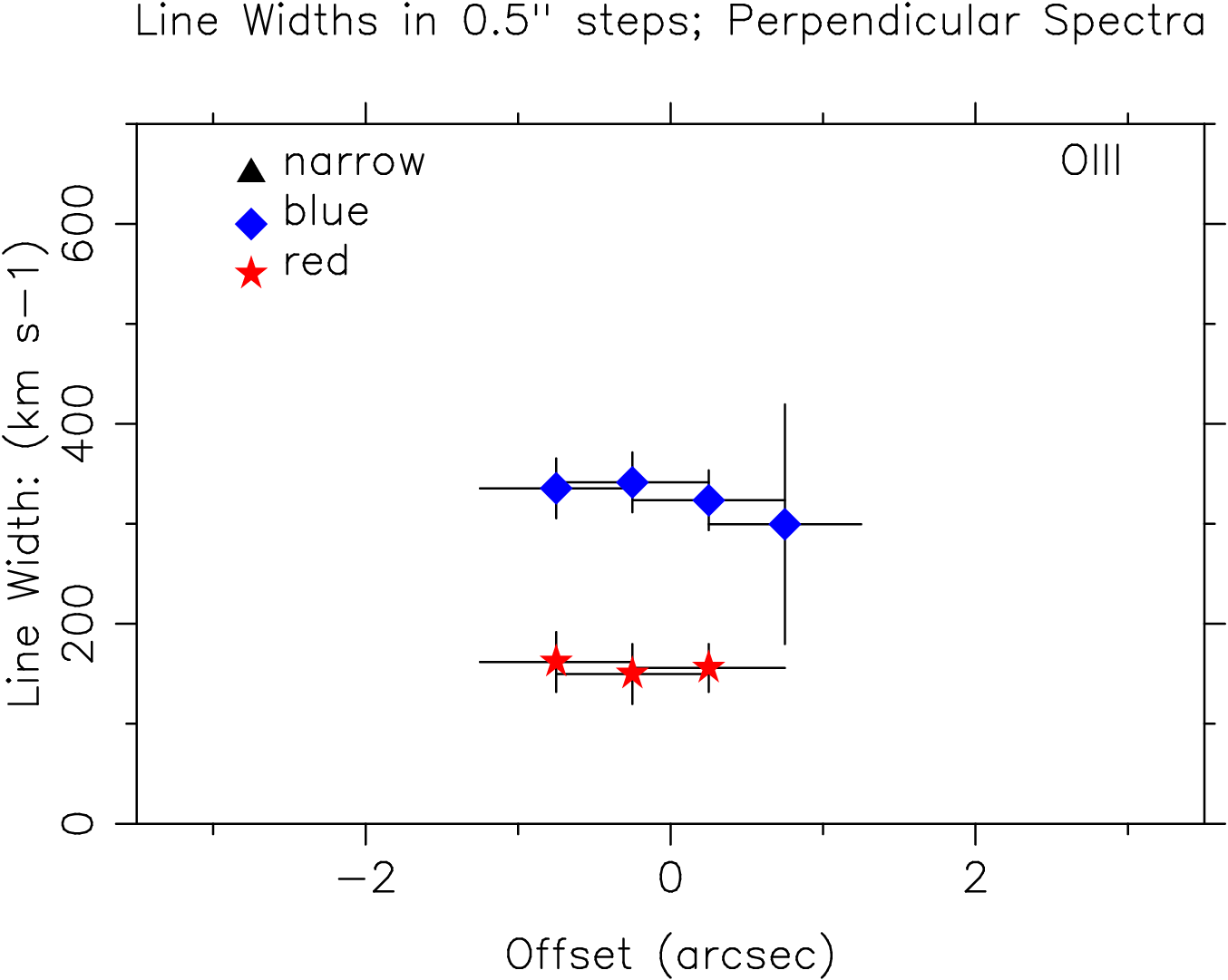}}\\
\resizebox{\hsize}{!}
{\includegraphics[angle=-90]{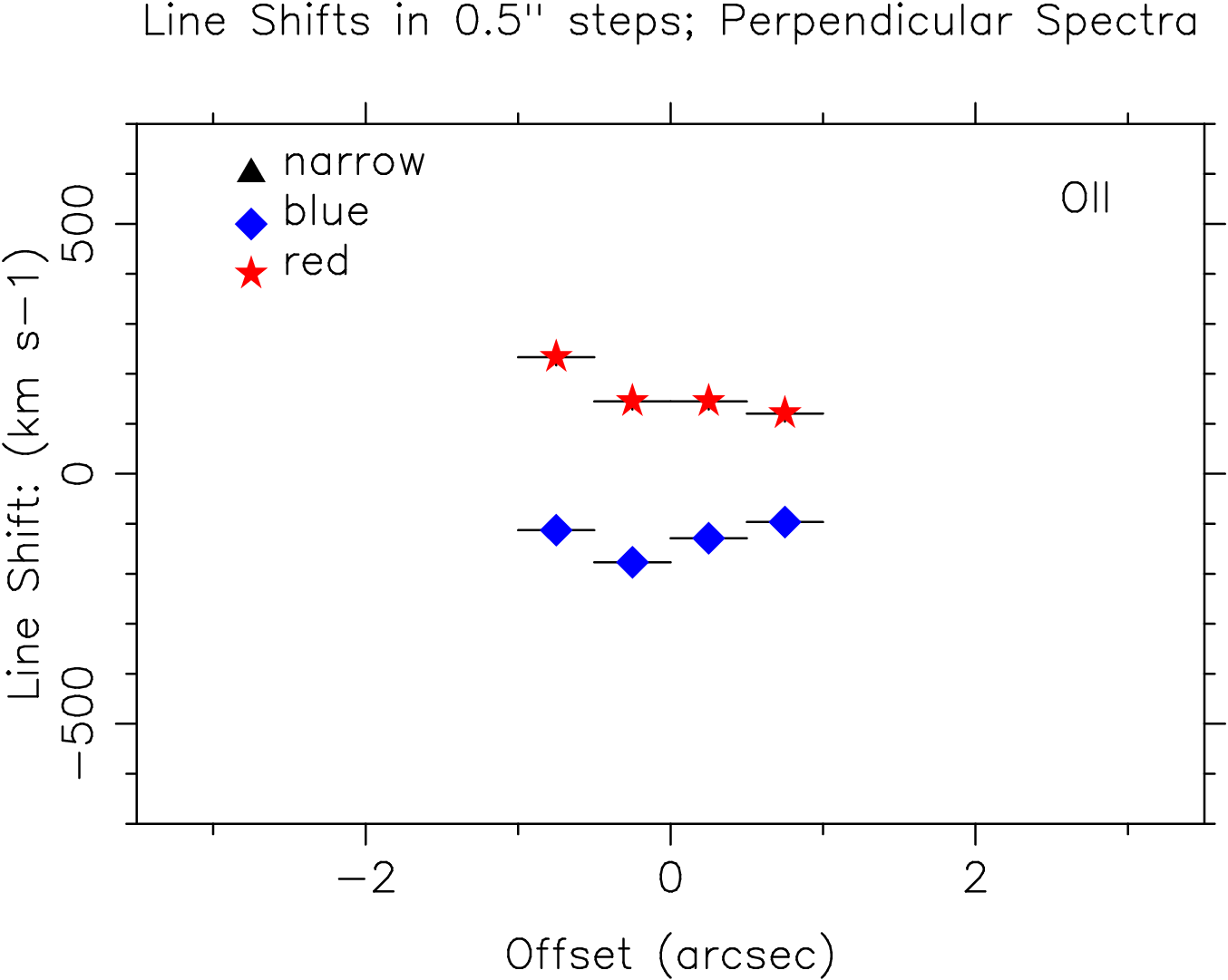}
\includegraphics[angle=-90]{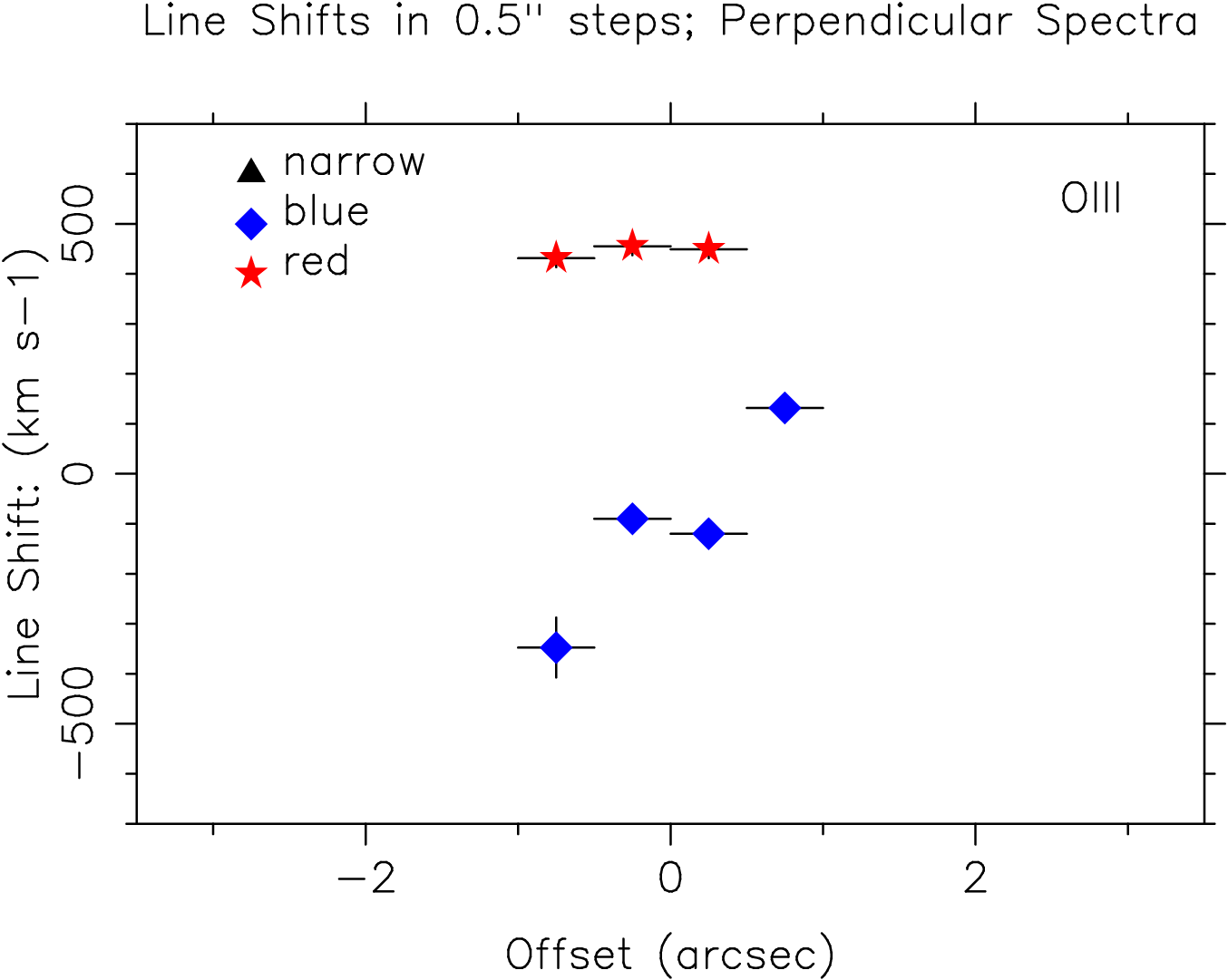}}

\caption{Gas kinematics for 9C J1503+4528.  The changing line fluxes
  (top), widths (centre) and offsets (bottom) with position are
  displayed for the \oo\ (left) and \o3\ (right) emission lines,
  extracted and fitted in 0.5$^{\prime\prime}$ steps from the perpendicular
  spectrum. Up to three emission line components are modelled at each
  position.  Negative offsets are towards the SW, and positive offsets
  towards the NE.}
\label{K2}
\end{figure}

We also present the results of a similar analysis of the spectrum
aligned perpendicular to the radio source axis (Fig.~\ref{K2}).  As
the line emission in the perpendicular spectrum is much less
extensive, larger extracted spectra of 1$^{\prime\prime}$ in width
have been modelled to boost the signal-to-noise.  The first main
feature of note is the lack of an extended narrow component.  This is
strong evidence that the source of the ionizing photons for the
extended narrow component is the ionization cone of the AGN, which
does not intersect with the perpendicular spectrum except in its
centre.

In the central regions of the perpendicular spectrum, broad emission
components are observed.  These are very similar to those in the
parallel spectrum, with the exception that there is no clear tendency
for the blue- or red-shifted emission to preferentially lie on one
side of the galaxy or the other.  This suggests that outflows along
the radio axis are a plausible option for explaining the gas
kinematics, as an alternative to simple rotation.  The greater
luminosity and extent of the material to the SE side of the galaxy,
together with the blueshifted broader component, suggests that the SE
radio lobe would be orientated towards the observer; given the
observed radio asymmetries (section 2.1) and quiescent extended 
emission line component at larger radii the outflow scenario is
indeed highly plausible.

\begin{figure*}
\begin{center}
\vspace{6.90 in}
\includegraphics{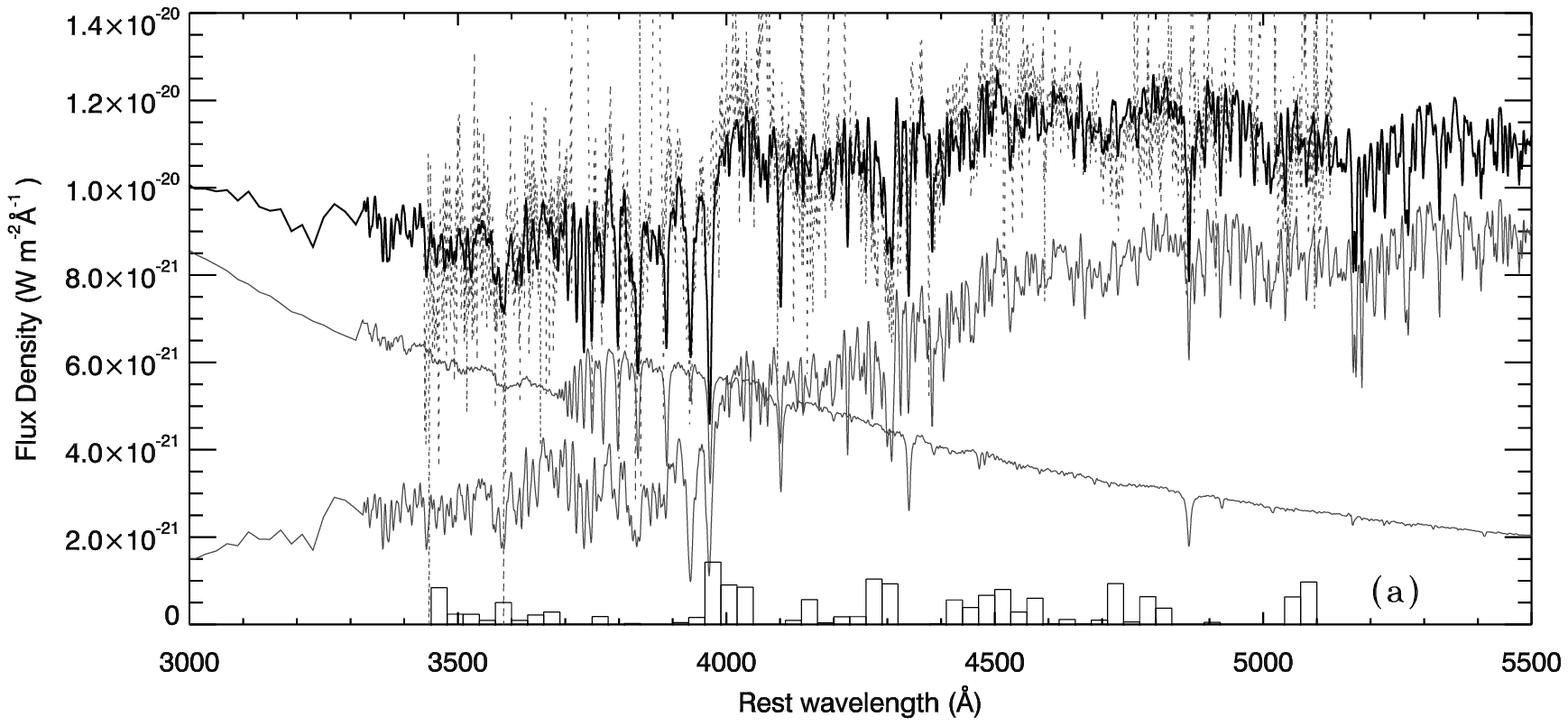}
\includegraphics{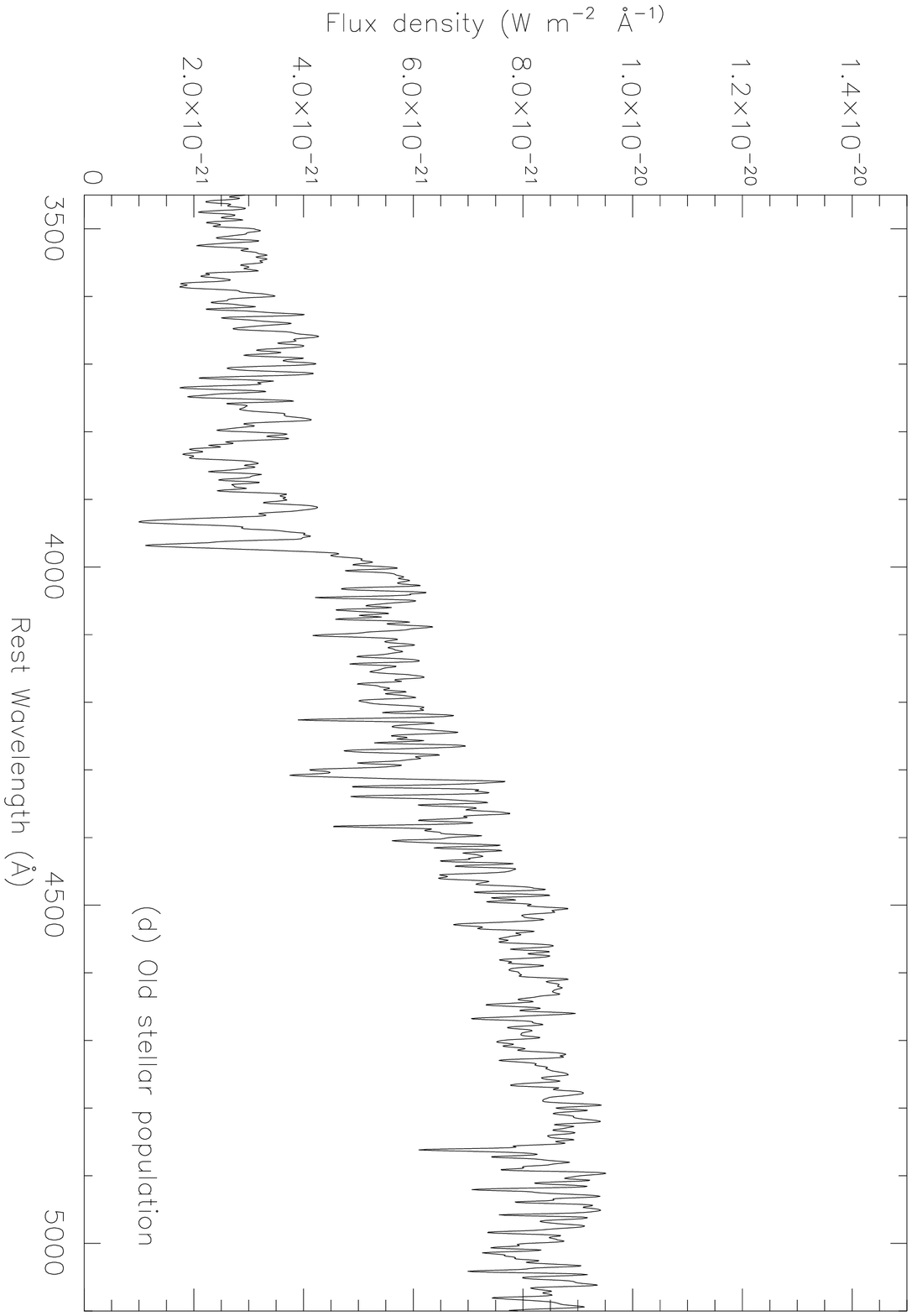}
\includegraphics{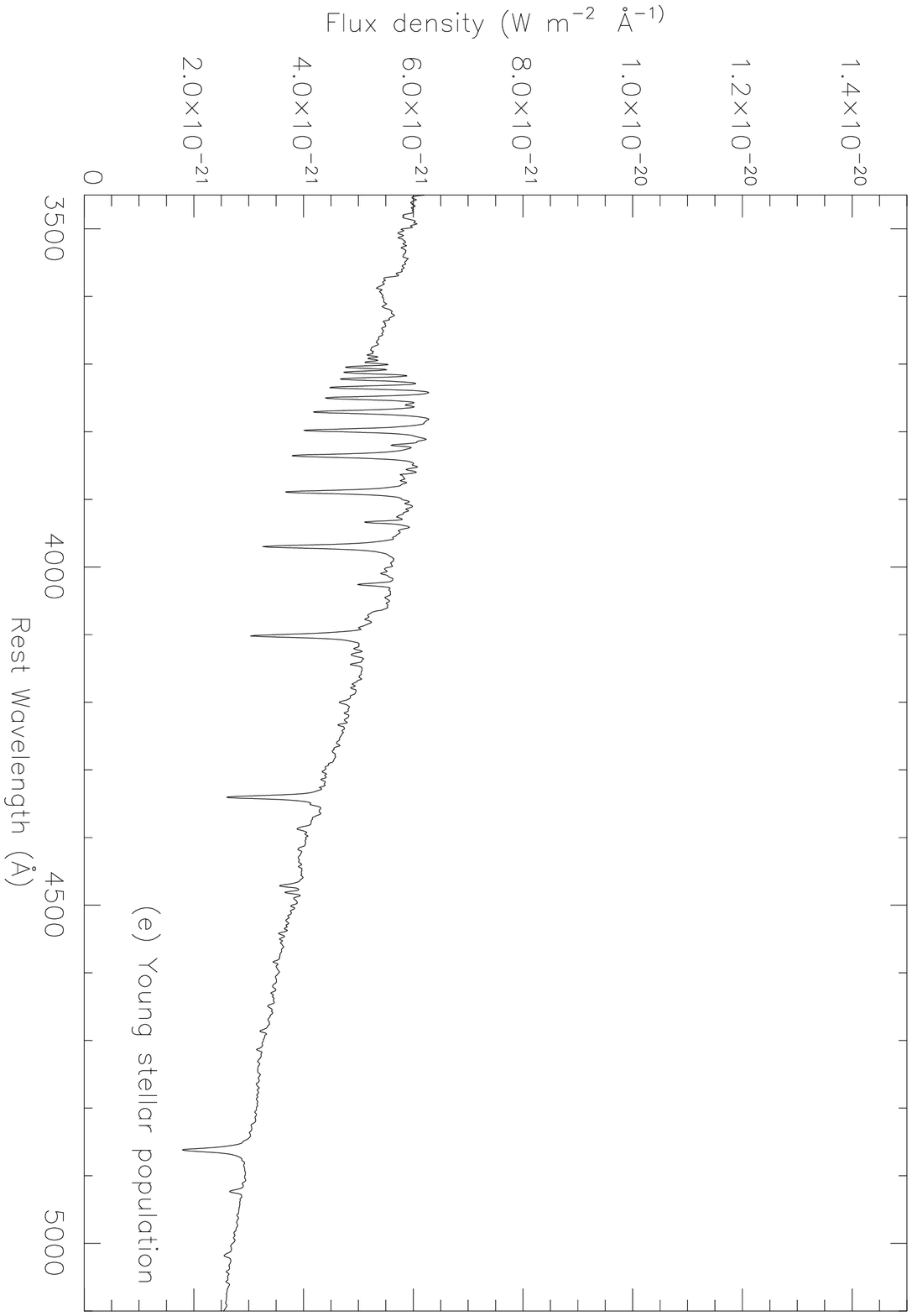}
\includegraphics{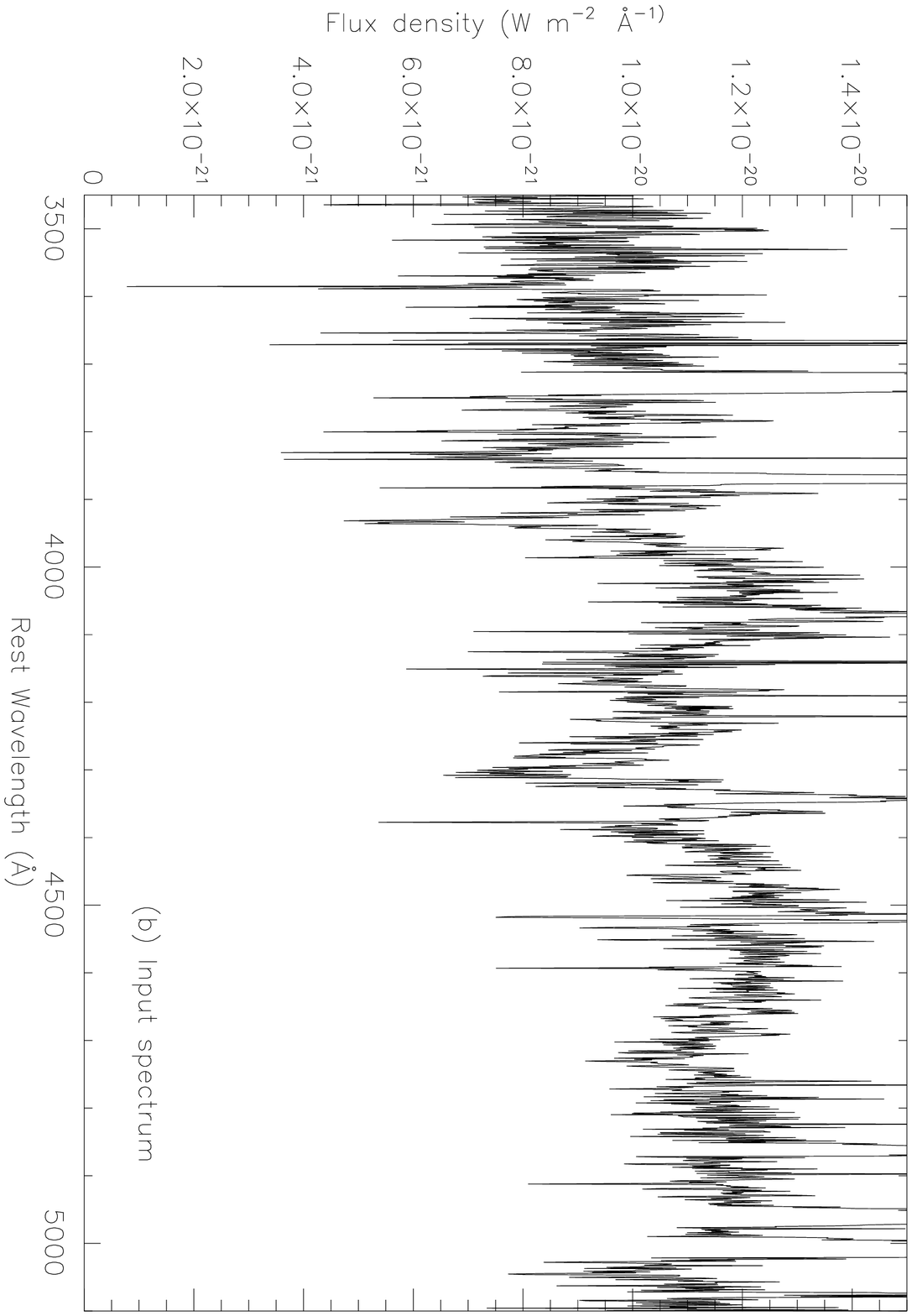}
\includegraphics{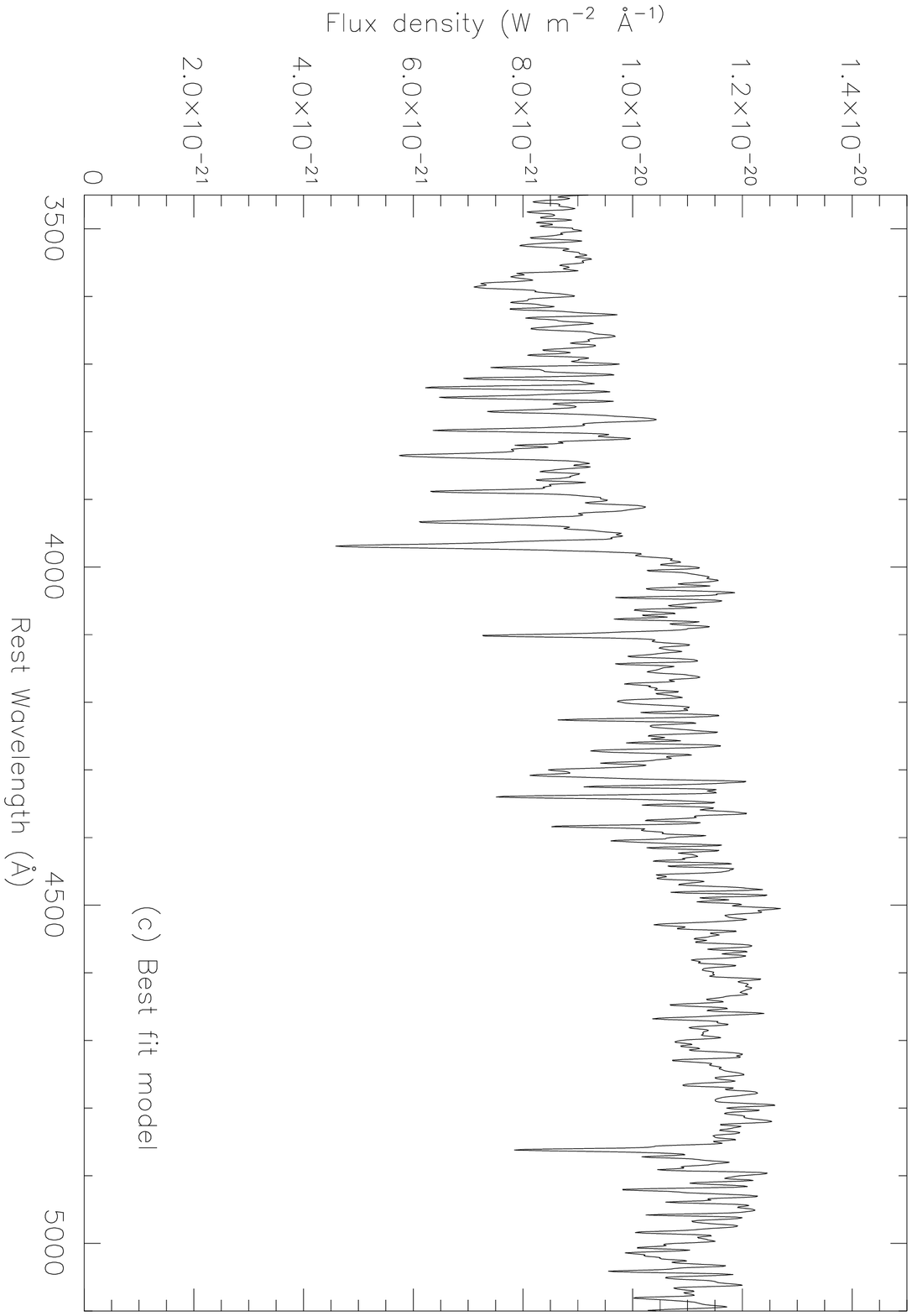}
\end{center}
\caption{Continuum fits to the central 2\arcsec of the combined
  spectrum (Fig.~\ref{nicespec}).  The top frame (a) displays the observed
  spectrum (dotted line), the best-fit model (dark solid) and the two component (old and young)
  SEDs which make   up the best-fit model.  Also
  displayed in frame (a) are the model residuals for each of the bins used in the fitting
  procedure.  For clarity, we
  also display the observed data (frame (b), centre left), the
  best-fit model (frame (c), lower left), the old stellar population
  component (a 6 Gyr old stellar population, frame (d), centre right) and the young stellar population
  component (a 5 million year old stellar population, frame (e), lower right) as separate spectra.
\label{confit}}
\end{figure*}
\section{Continuum emission}

The continuum emission of 9C1503+4528 is remarkably blue.  Together
with the clear presence of Balmer absorption lines and the young age
of the radio source this suggests a significant young stellar
population.  If the triggering of radio source activity is connected
with the star formation, the properties of any young stellar
population formed as part of that process can shed light on the exact
mechanisms and timescales involved.  We have therefore carried out
continuum modelling of our spectra, with the aim of placing limits on
both the mass and age of any young stellar population present in this
galaxy.

For the continuum modelling, we combined both the parallel and
perpendicular 2-d frames and extracted a 1-d spectrum from the central
2$^{\prime\prime}$ (Fig.~\ref{nicespec}).  

The first stage was to remove the contribution of nebular continuum
emission (Dickson et al 1995). Although subject to stellar absorption
features, the Balmer line strengths are in general consistent with the
predictions of case B recombination, and internal reddening is
unlikely to be a major issue.  The H$\beta$ flux can be well modelled,
and was used to determine the contribution from the various nebular
continuum emission processes (free-free emission, bound-free emission,
two-photon emission and the Balmer forest pseudo-continuum), assuming an electron temperature of
15000K. (Checks with alternative temperatures, allowing
for the error on our evaluation of the H$\beta$ flux, do not produce
significantly different results.)
The corrected spectrum was binned in blocks of $\sim 30$\AA, excluding any
regions where emission lines were present.  

%Reduced $\chi^2$ fitting of our extracted continuum
%spectrum was carried out using 

To fit the remaining stellar continuum we used a combination of young
and old stellar population SEDs from the models of Bruzual \&
Charlot (2003). 
A Chabrier initial mass function and solar metallicity were
selected, and the stellar populations were assumed to form in an
instantaneous burst. We consider stellar population ages ranging from $< 10^{5}$ years up to $9
\times 10^{9}$ years (the expected age of the old stellar population
in a $z \sim 0.5$ galaxy, assuming a formation redshift of $z \sim
10-20$).  Reduced $\chi^2$ fitting was used to determine
the relative proportions of old/young stars for all combinations of
SED models.

Figure \ref{confit} displays the results of our best-fit model.  This
consists of an old (6 Gyr) stellar population, plus a $5 \times 10^6$
year young stellar population accounting for 0.065\% of the total
stellar mass, and roughly 70\% of the total continuum flux at a
wavelength of $\sim 3600$\AA.   Our modelling implies that the total
stellar mass in the central region of the galaxy probed by our spectra
is approximately $10^10 M_\odot$, whereas the overall stellar mass of
the galaxy (based on our $K-$band modelling in section 6) is closer to
$10^11 M_\odot$. 
The largest residuals lie near
the Ca absorption feature at $\sim 3970$\AA, which is subject to
infilling by line emission at 3967\AA\ due to H$\epsilon$ +
[Ne\textsc{iii}].

\begin{figure}
\vspace{2.20 in}
\begin{center}
\includegraphics{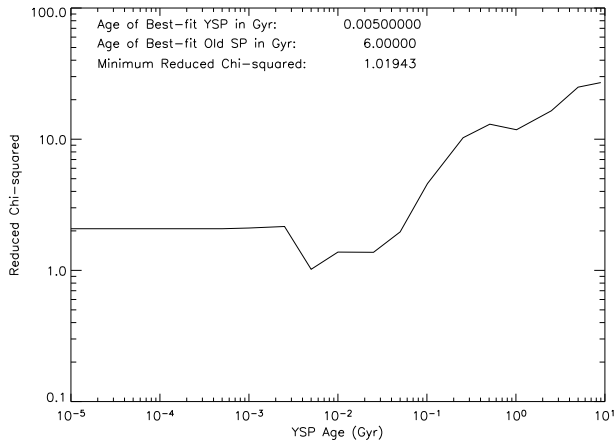}
\end{center}
\caption{Reduced $\chi^2$ of best-fit model vs. young stellar
  population component age for the continuum fitting of 9C
  J1503+4528.  The best-fit model includes a YSP component with an age
  of $5 \times 10^6$ years, accounting for 0.065\% of the total
  stellar mass of the host galaxy, and roughly 70\% of the observed
  continuum flux at a wavelength of $\sim 3600$\AA. The age of the
  majority old stellar population in the best-fit model is
  approximately 6Gyr.
\label{confit_chi}}
\end{figure}

A plot of best-fit reduced $\chi^2$ as a function of young stellar
population (hereafter YSP) age is
displayed in Fig.~\ref{confit_chi}.  A contribution from a young
stellar population is very much favoured over an old stellar
population alone.
Fits obtained using different ages for the majority old stellar
population (ranging from 5 to 9 Gyr in age) do not lead to any
significant variation in the resulting best-fit YSP.

We must also consider the potential presence of a power-law ionizing
continuum associated with the AGN. If the presence of a power-law
component is simply added to our modelling, allowing its flux and shape to
vary freely, the resulting best-fit model generally prefers an 
unrealistically strong power-law flux, and it is therefore necessary
to place constraints on the power law contribution in order to
make a meaningful assessment of its effects.
To determine the balance between YSP age/mass and power-law
contribution, we repeated our fitting procedure using fixed power-law
contributions ranging from 0\% to 50\% of the total flux at a
wavelength of $\lambda \sim 3600$\AA.  The spectral index of the power
law (defined as $F_{\lambda} \propto \lambda^{\alpha}$) is constrained
to be one of three values, $\alpha = -1$, $-2$ or $-3$.  The fitting
procedure was then repeated for all combinations of 
stellar populations and power-law parameters.  The resulting best-fit YSP parameters (age, mass, reduced
$\chi^2$) for each of the different power law contributions
considered are displayed
in Appendix B.

With the exception of the fits with the very highest power-law
contribution ($> 25\%$), the
best-fit YSP mass and age remain relatively unchanged. 
The similar strengths of the narrow line emission
components either side of the galaxy centroid in the parallel spectra
(suggesting that this source is observed with the ionization cone
aligned in a direction nearly perpendicular to the line of sight)
also imply that a two-component stellar population provides a more than adequate explanation of the
observed continuum emission, and that no power-law is required.

\section{Implications for source triggering}

\subsection{Morphological properties of the host galaxy}

Given the spectroscopic evidence for a young stellar population, we
examined the galaxy morphology for evidence for a merger or other
interaction which may 
have caused the starburst.  Our calculated 0.06\% 5\,Myr YSP would be
expected to contribute only 0.02 magnitudes to the $K$-band flux, and
so the $K$-band image will be dominated by the old stellar populations
of the host galaxy and any merging companion or tidal tails.

We fitted the galaxy morphology using the methods detailed in Inskip
et al.  (2005), which can be briefly summarised as
follows.  The point spread function for the UFTI field is determined
from unsaturated stars within the mosaic field; we extract their 2-d
profile, normalise to unit flux and take the average profile.
S\'{e}rsic profile (S\'{e}rsic 1968) galaxy models were then convolved with the point
spread function and fitted to the surface profile of the galaxy,
using available least squares minimisation IDL
routines\footnote{\textsc{mp2dfunfit.pro}, part of Craig Markwardt's
\textsc{mpfit} non-linear least squares curve fitting package
available via http://astrog.physics.wisc.edu/
$\sim$craigm/idl/fitting.html.}.  Free parameters are the galaxy flux,
centroid, effective radius, S\'{e}rsic index and fractional nuclear
point source contribution. Galaxy ellipticity was also allowed to
vary, but in the case of this object a circular model with $\epsilon =
0$ was preferred.

\begin{figure}
\vspace{3.42 in}
\begin{center}
\includegraphics{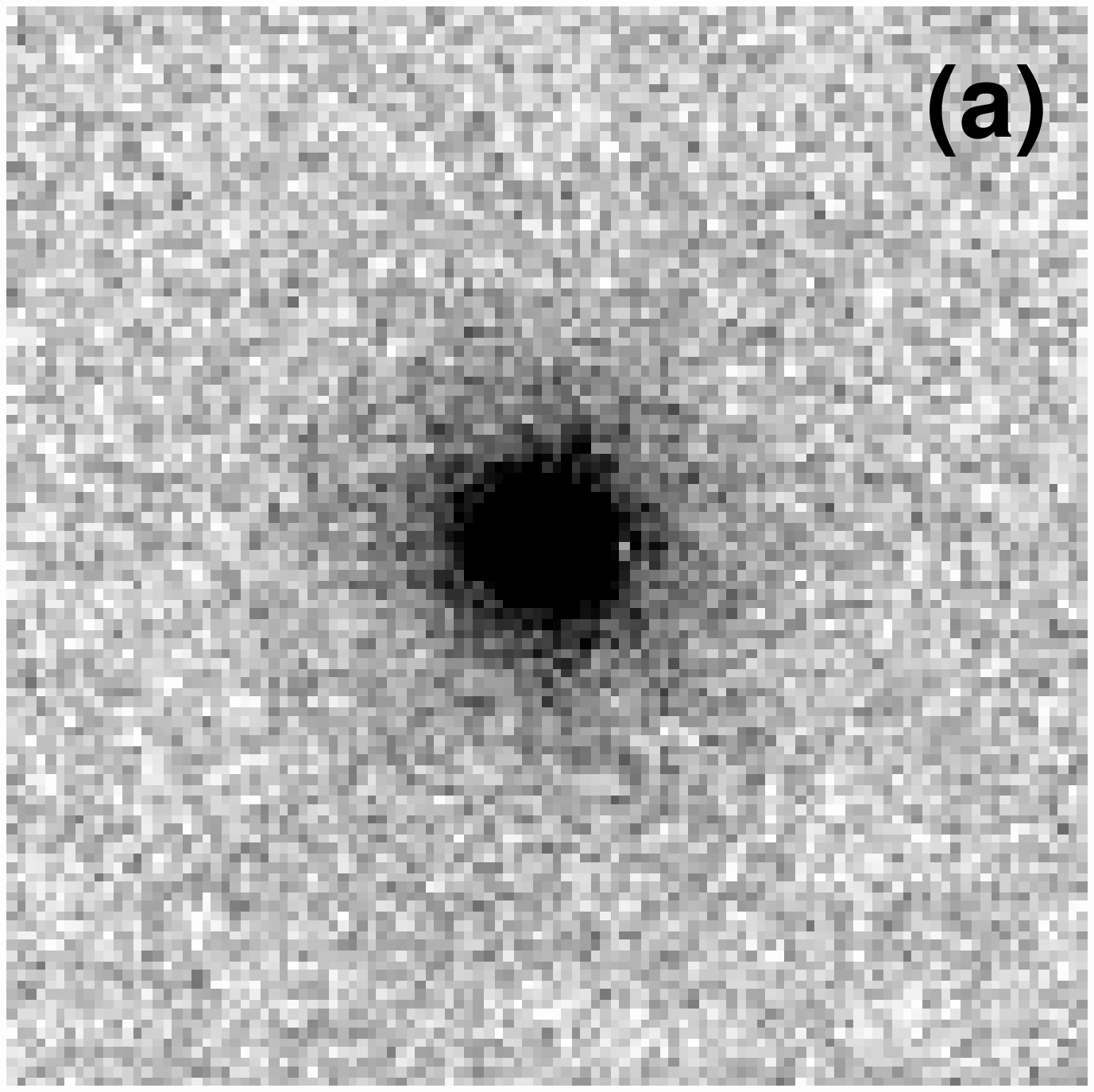}
\includegraphics{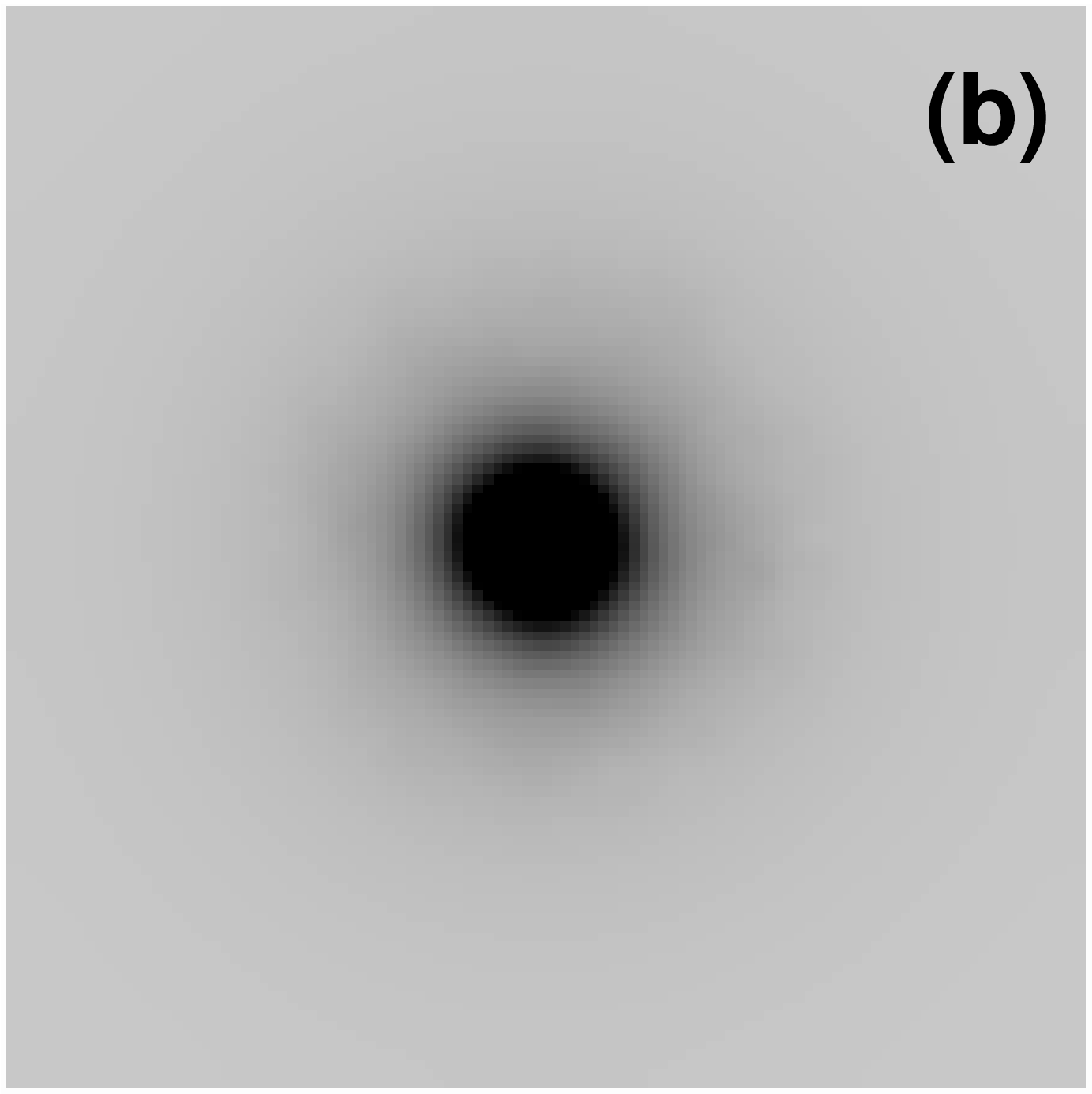}
\includegraphics{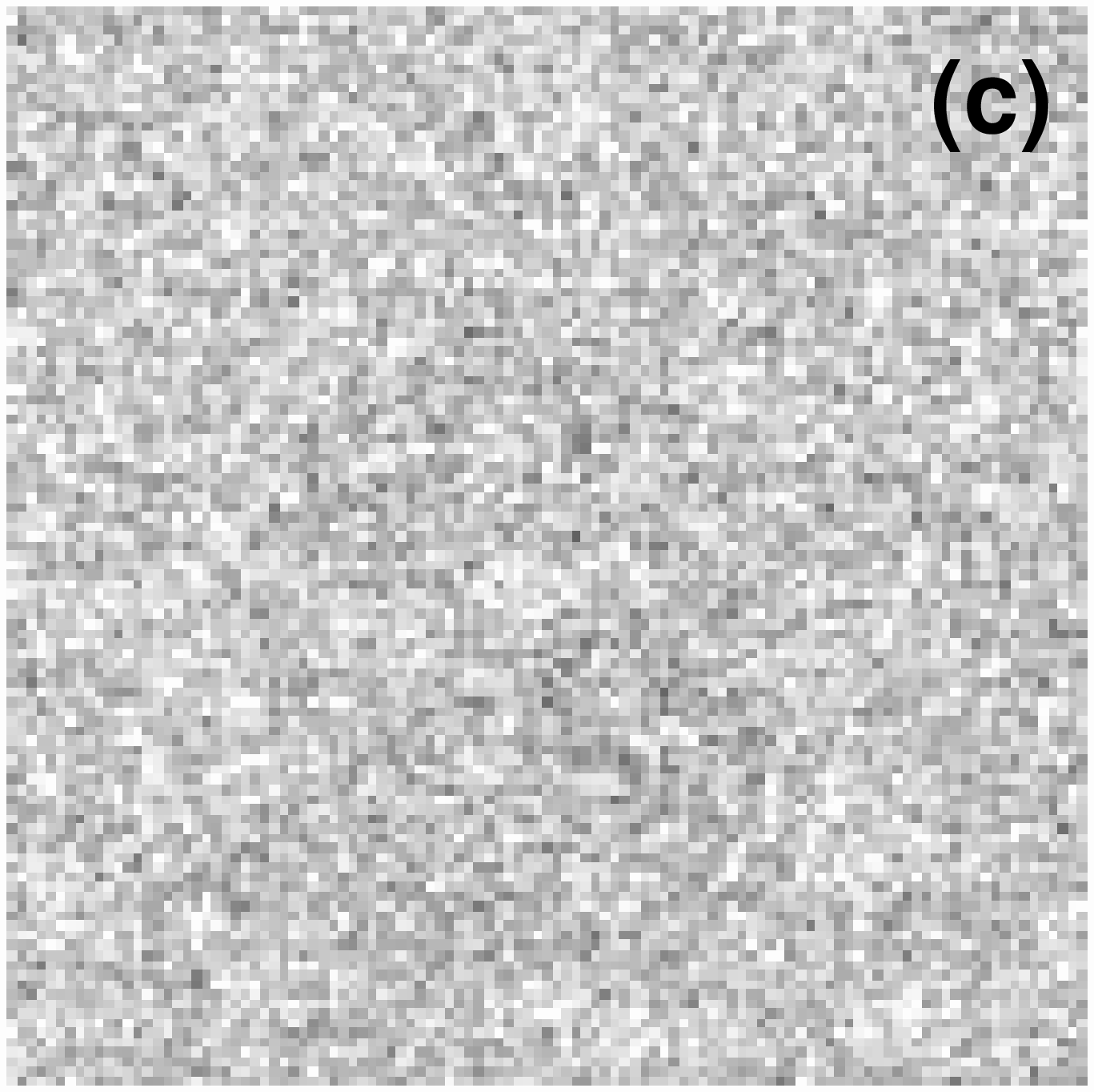}
\includegraphics{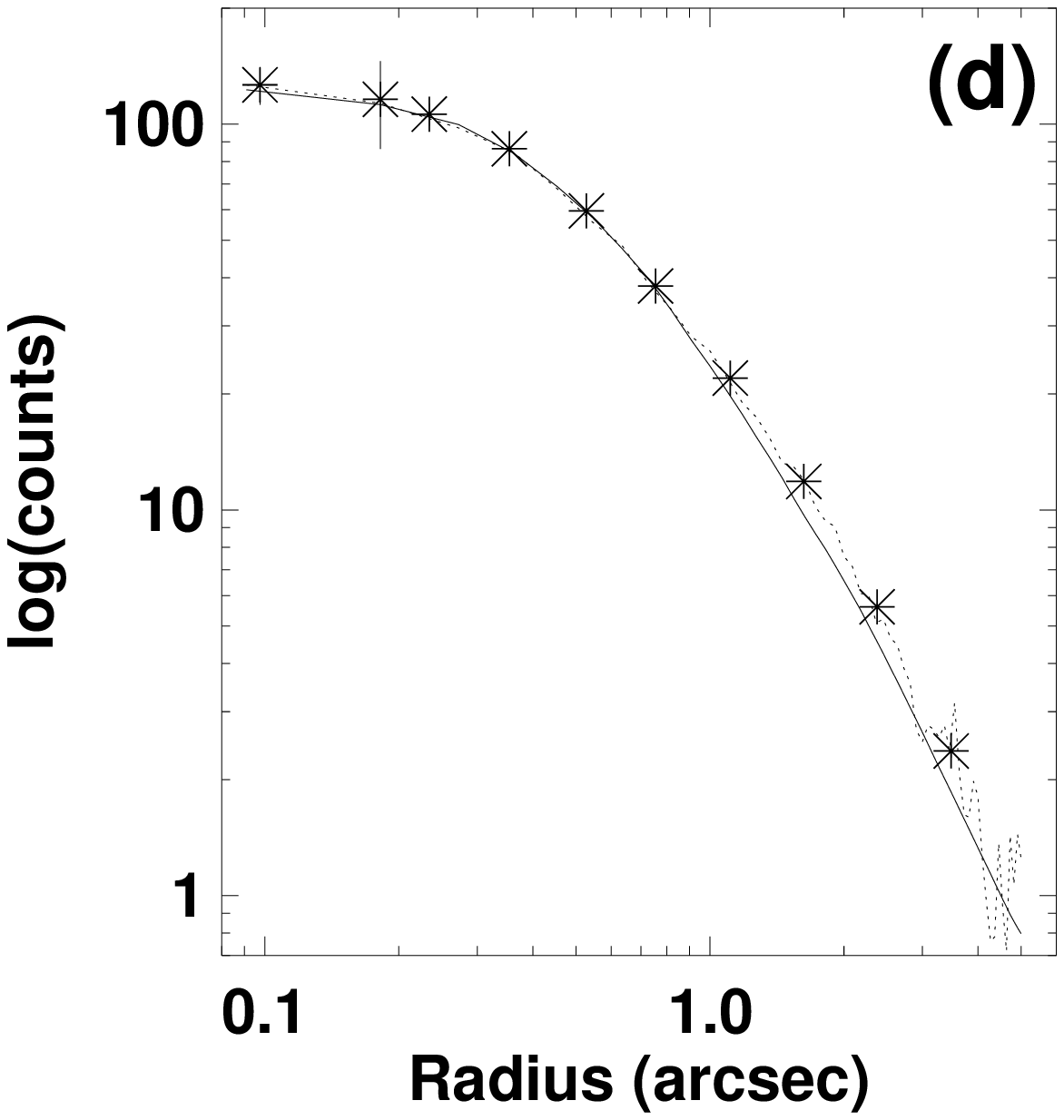}
\end{center}
\caption{Host galaxy morphology fits for 9C J1503+4528.  The $K-$band
   galaxy image is presented in frame (a), with the model galaxy given
   in frame (b) and the residuals after subtracting the best fitting model from
   the data displayed in frame (c).  Frame (d) illustrates a 1-d cut
   through the data (dotted line and binned points) with the best-fit model 
(solid line)
   overlaid.
\label{Fig: morph}}
\end{figure}

From the $K$-band data, our resulting best-fit
model, residuals and 1-d radial profile are displayed in
Fig.~\ref{Fig: morph}. We find that the host galaxy appears to have a
simple de Vaucouleurs elliptical morphology, with a preferred
S\'{e}rsic index of $4.0 \pm 0.2$, and a best-fit ($\chi^2 = 1.03$)
effective radius of $r_{\rm{eff}} = 1.70^{\prime\prime} \pm
0.15^{\prime\prime}$ (equivalent to $11.5 \pm 1.0$kpc at
$z=0.521$). We find no evidence for any nuclear point source
contribution at $K$ band (preferred point source percentage $<1\%$).

The $r$-band data are too noisy to make an accurate fit to the data;
the effective radius is constrained only to lie between
1.3$^{\prime\prime} \pm 0.5^{\prime\prime}$ and 1.6$^{\prime\prime}
\pm 0.5^{\prime\prime}$. This is, however, consistent with the
$K$-band fit.  
Most importantly, we find no significant residual emission which would indicate
the presence of a close companion galaxy or tidal tails.

Any galaxy interaction/merger activity must therefore either have been
very minor in nature, or have occurred sufficiently long ago that the
old stellar population of the host galaxy has returned to an
undisturbed morphology. In the latter case, the interaction must have
taken place several dynamical timescales before the present
observations. We estimate the dynamical timescale $T_{dyn}$ as $r_{\rm{eff}}/
\sigma$, where $\sigma$ is the velocity dispersion of the galaxy,
which in turn we estimate by taking the FWHM of the narrow component
of the emission lines ($\sim 200$ km s$^{-1}$).  This gives $T_{dyn}
\sim $ 60 Myr, so any previous interaction must have occurred at a
time very much earlier that the formation of the YSP. We thus conclude
that the YSP and present AGN activity were caused by, at most, a minor
interaction.

\subsection{Evidence from the  extended emission-line region}

As with many other radio galaxies the properties of the extended
emission line region of 9C J1503+4528 vary substantially with spatial
position.  The changes in ionization state confirm our expectations
that shocks are important in the vicinity of the young radio source,
and that AGN photoionization dominates elsewhere.  The kinematic
properties of the emission line gas confirm this picture, with the
broader emission lines highlighting the disturbed kinematics of the
central regions, in contrast to the kinematically quiescent material
at larger radii.

The projected physical size of the EELR is $\approx 34$ kpc. Emission
line regions of this scale are by no means unusual, and are often
substantially larger in the case of FRII radio sources (for example,
$\sim$50--100 kpc for extended 3C and 6C sources, Best, R\"{o}ttgering
\& Longair 2000; Inskip et al 2002).  However, there are significant
contrasts between this EELR and those of more extended radio galaxies.

First, the EELR of 9C J1503+4528 has a very sharp cutoff, rather than
the more gradual decline in emission line luminosity observed in the
cases of other sources (e.g. Best, R\"{o}ttgering \& Longair 2000;
Inskip et al 2002). Second, and of great significance, is the fact
that the EELR is a factor of ten larger than the size of the
radio source. Unlike the case of the 3C and 6C sources, where the EELR
is significantly affected by the passage of the radio source, here we
have a case where the EELR could very well correspond to the limiting
distance reached by the ionization front of an AGN, illuminating the
region ahead of the radio source before it leaves the centre of the
host galaxy.

This factor of $\approx 10$ difference between radio source size and
EELR size has a significant implication. In combination with the
typical 0.1 c source expansion speed, we find it highly plausible that
the radio jets and AGN were triggered simultaneously, some $\sim 10^4$
years prior to our observations. 

\subsection{Spatial distribution of star formation}

\begin{figure}
\resizebox{\hsize}{!}
{\includegraphics[angle=90]{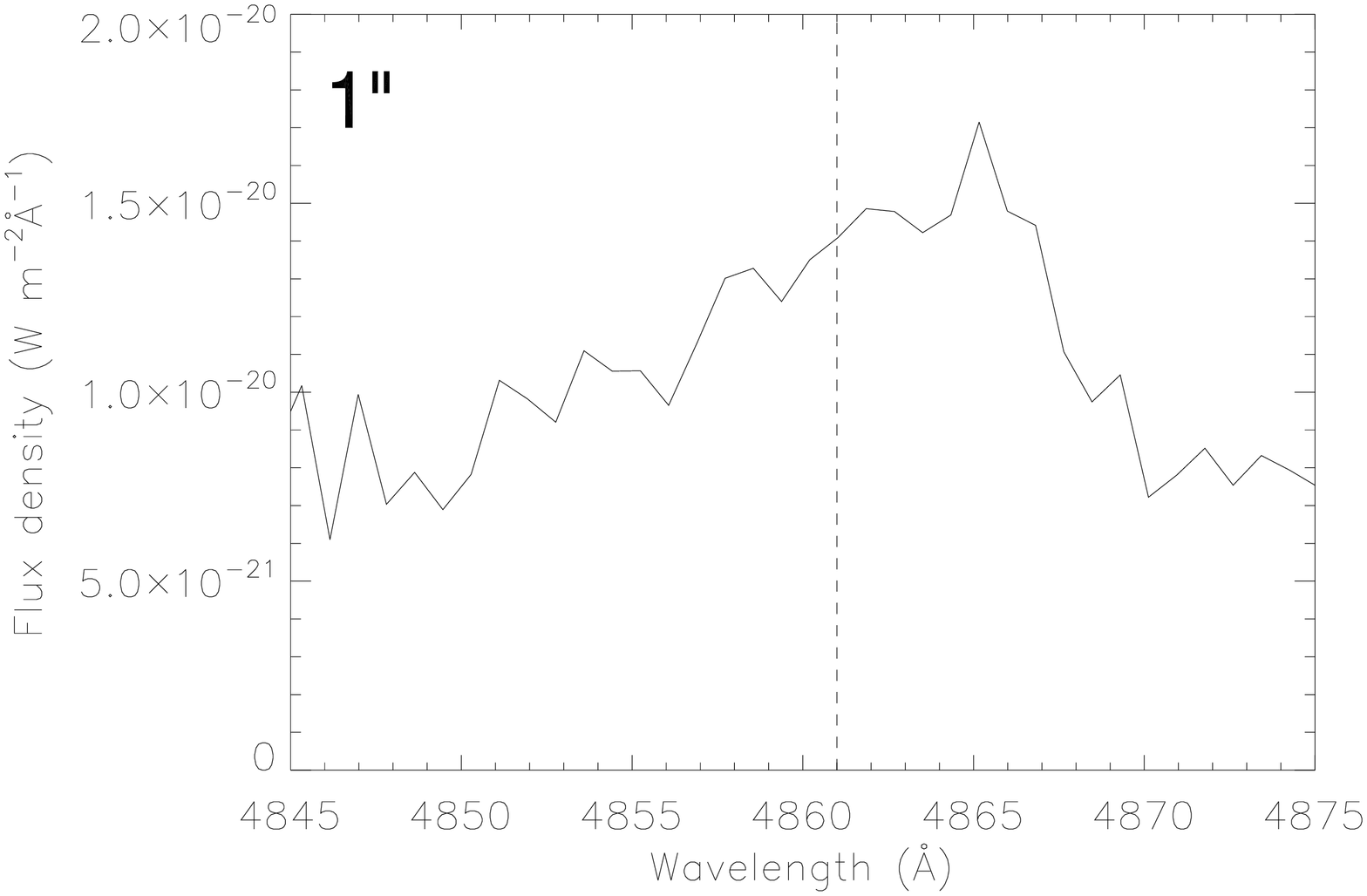}
\includegraphics[angle=90]{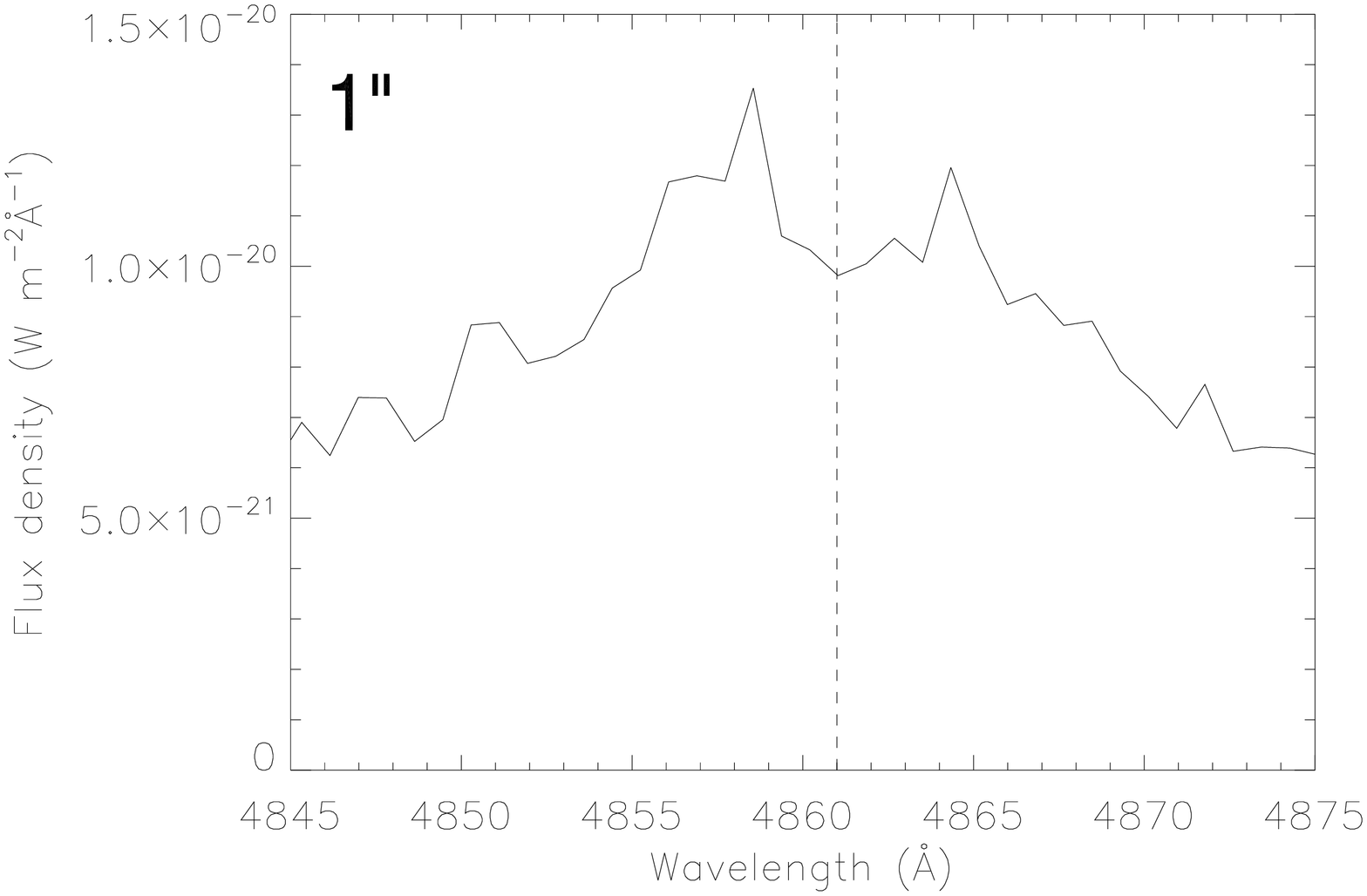}}\\
\resizebox{\hsize}{!}
{\includegraphics[angle=90]{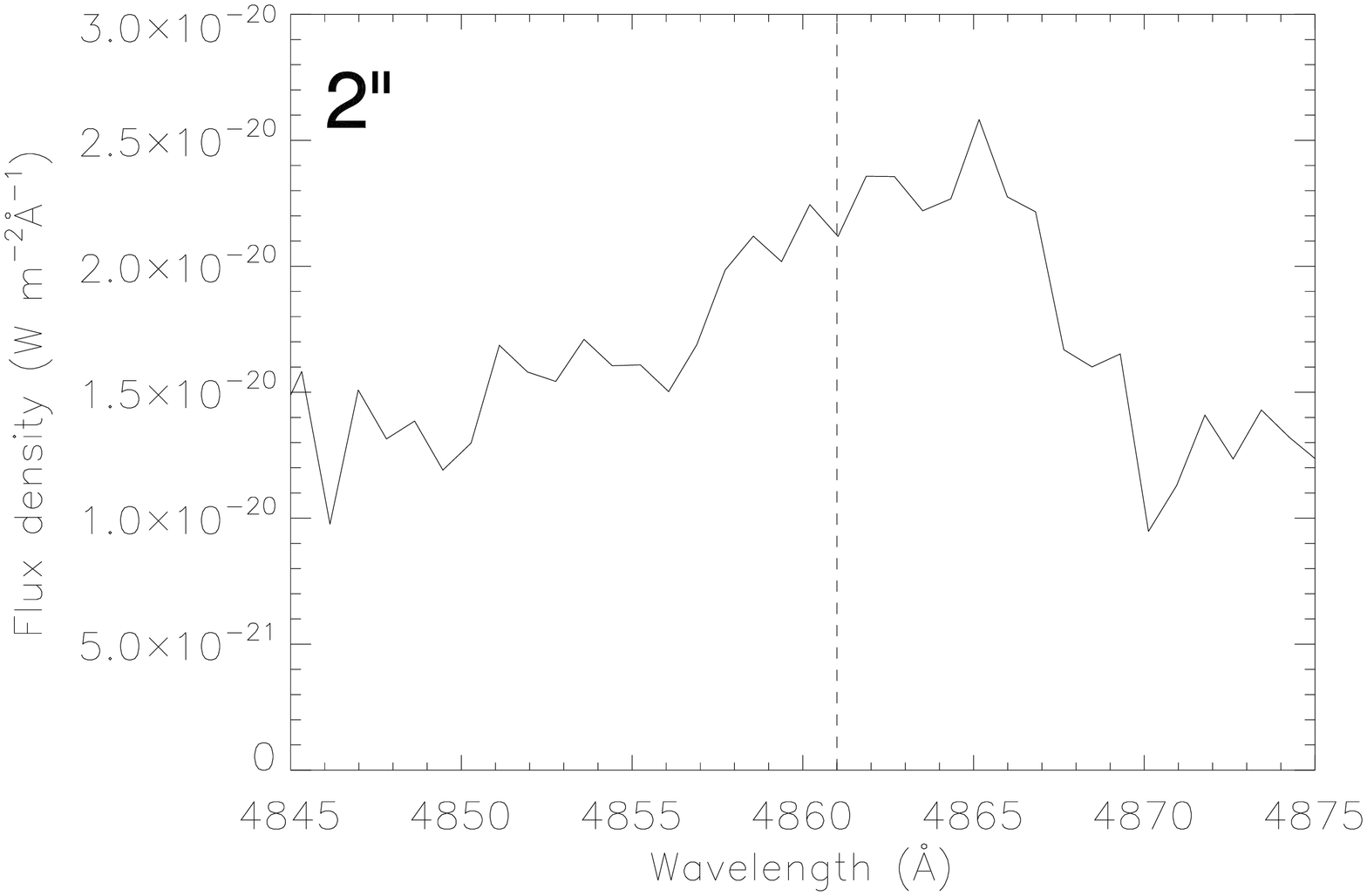}
\includegraphics[angle=90]{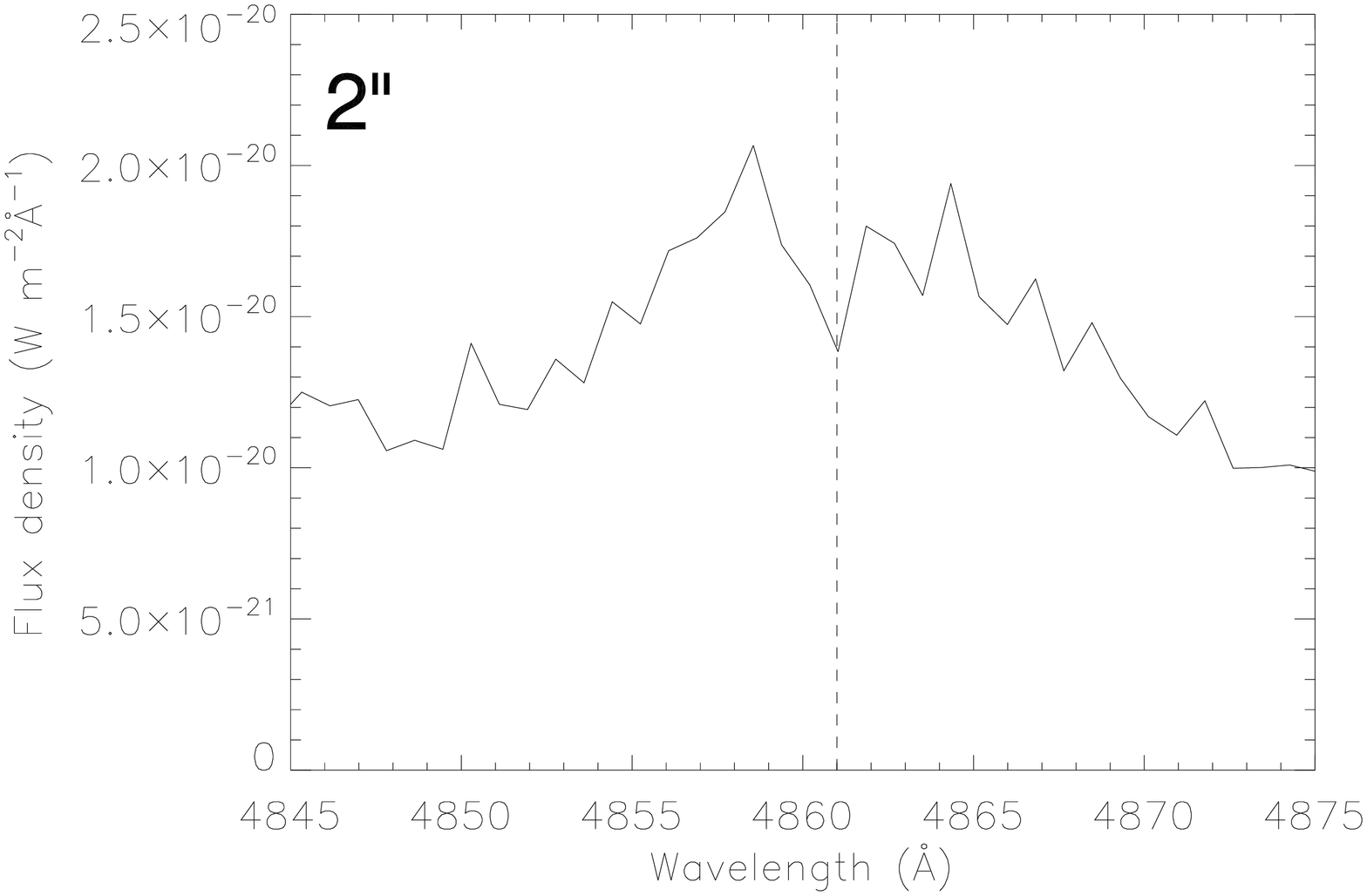}}\\
\resizebox{\hsize}{!}
{\includegraphics[angle=90]{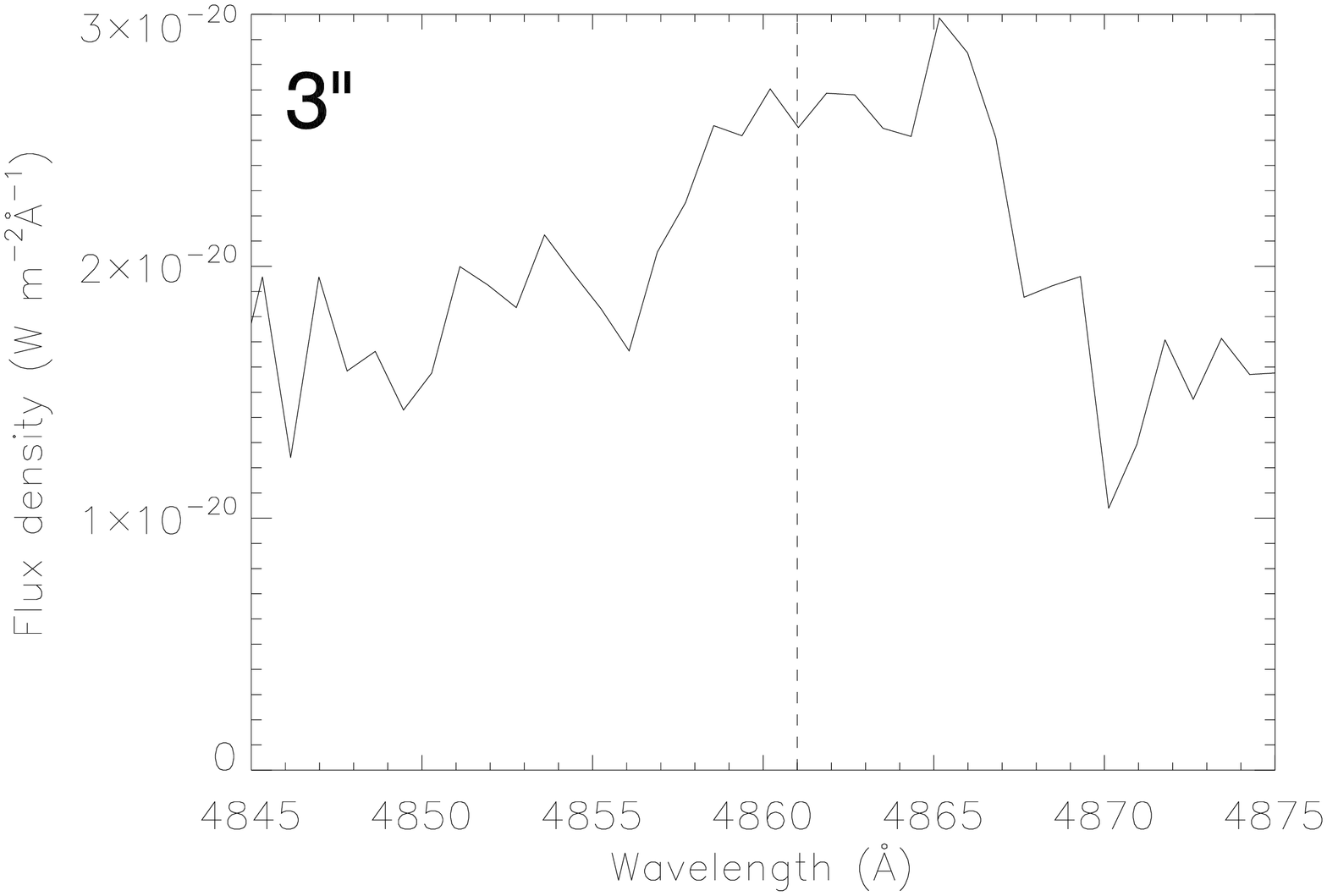}
\includegraphics[angle=90]{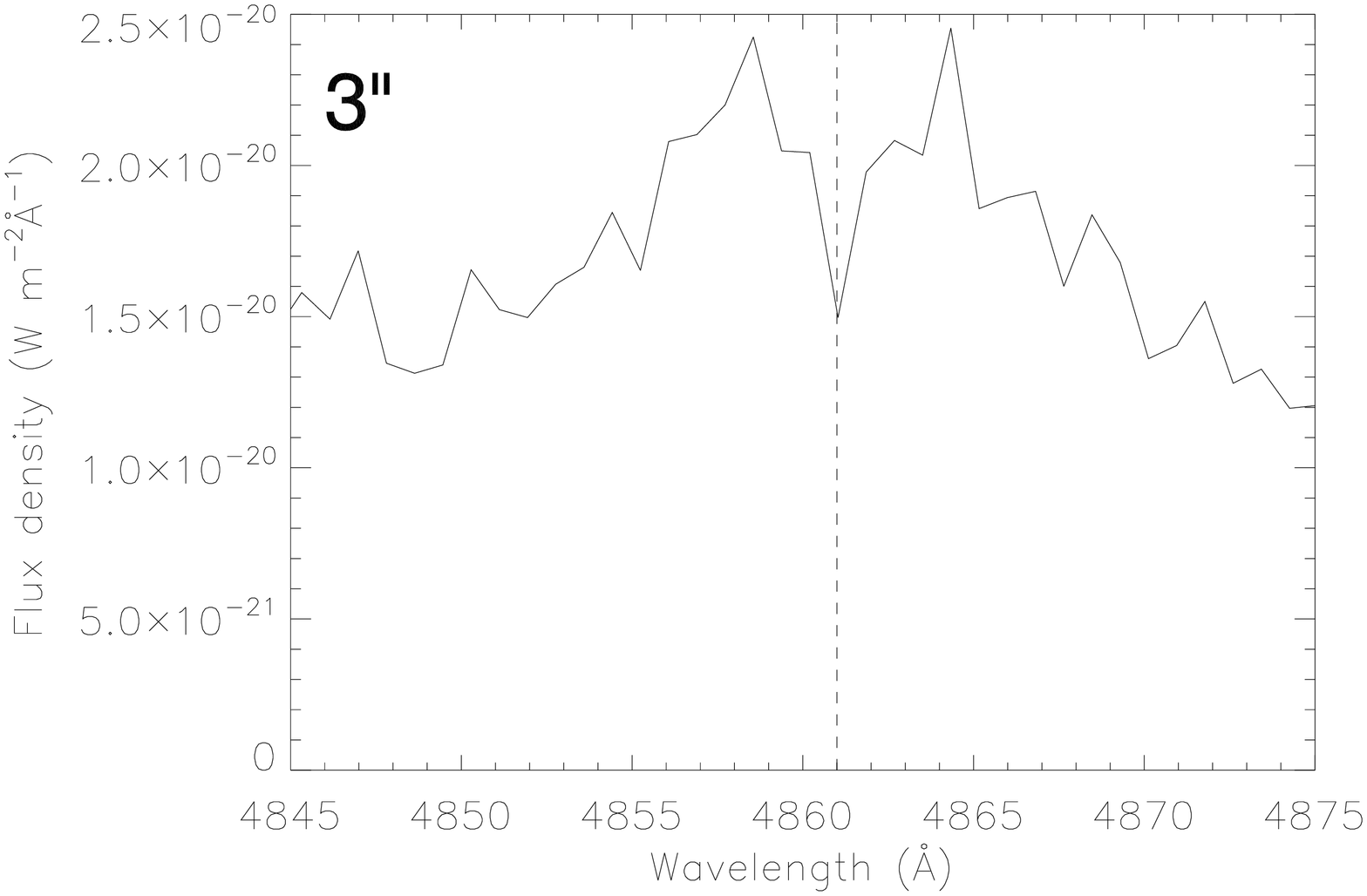}}\\
\resizebox{\hsize}{!}
{\includegraphics[angle=90]{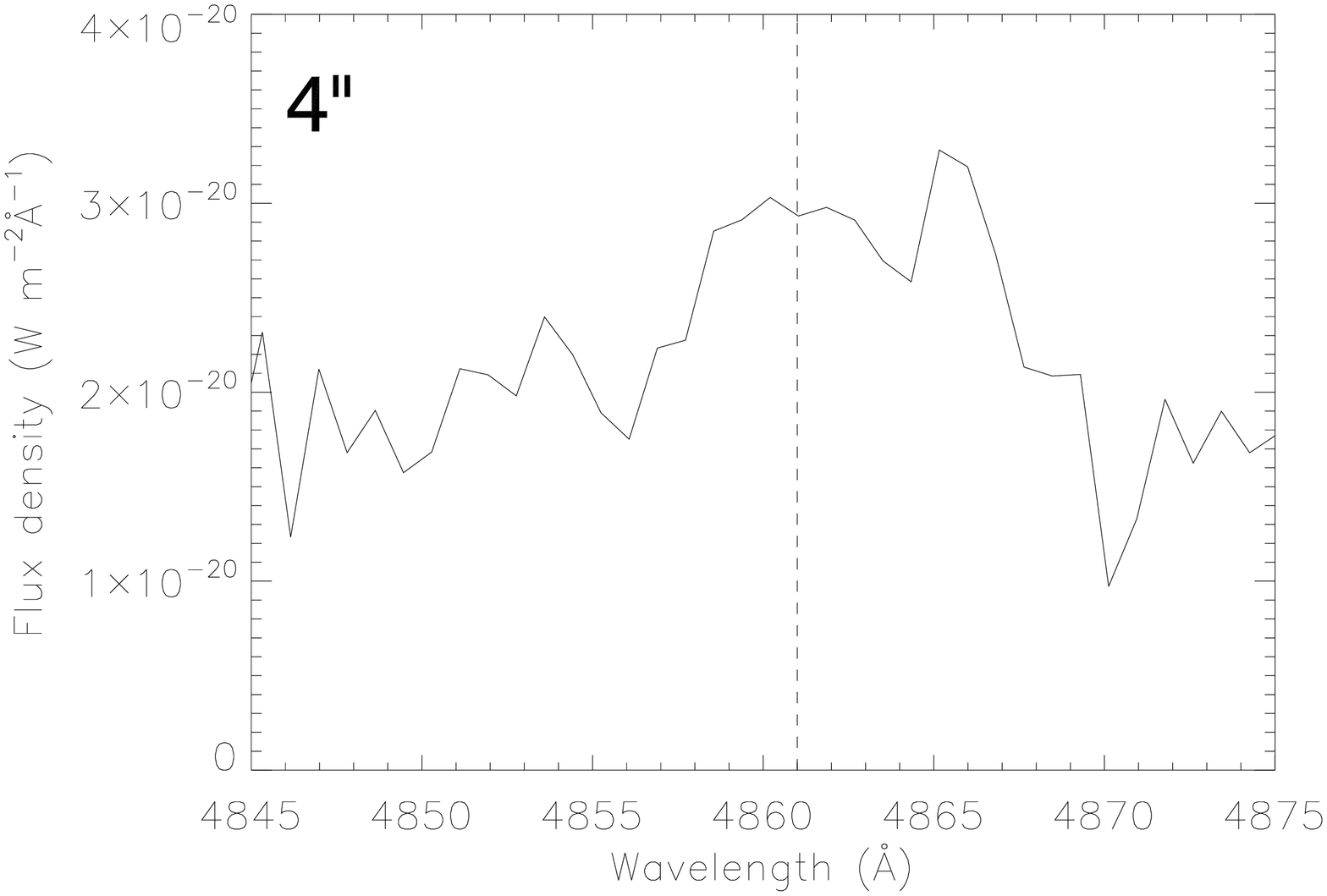}
\includegraphics[angle=90]{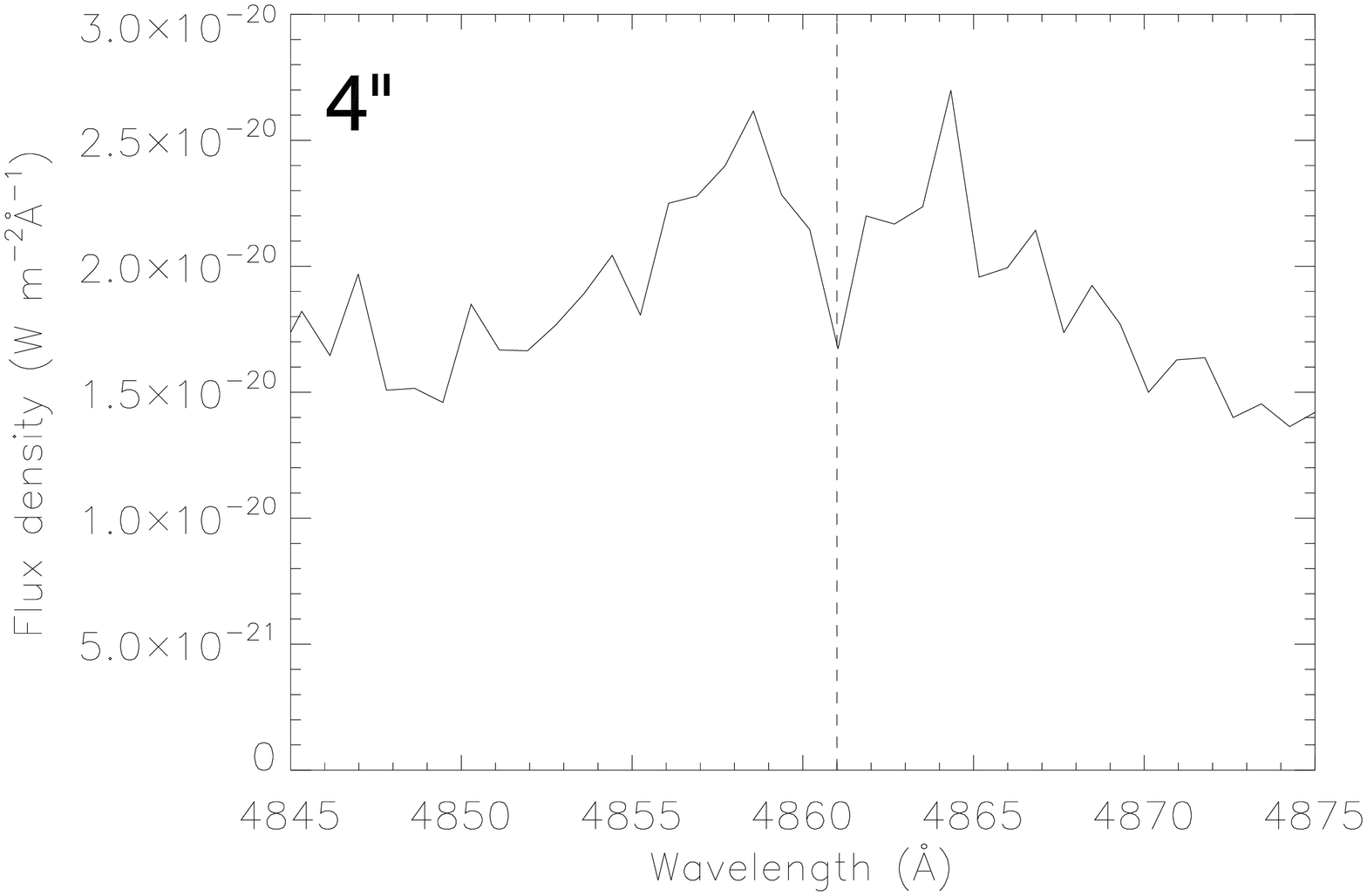}}\\
\resizebox{\hsize}{!}
{\includegraphics[angle=90]{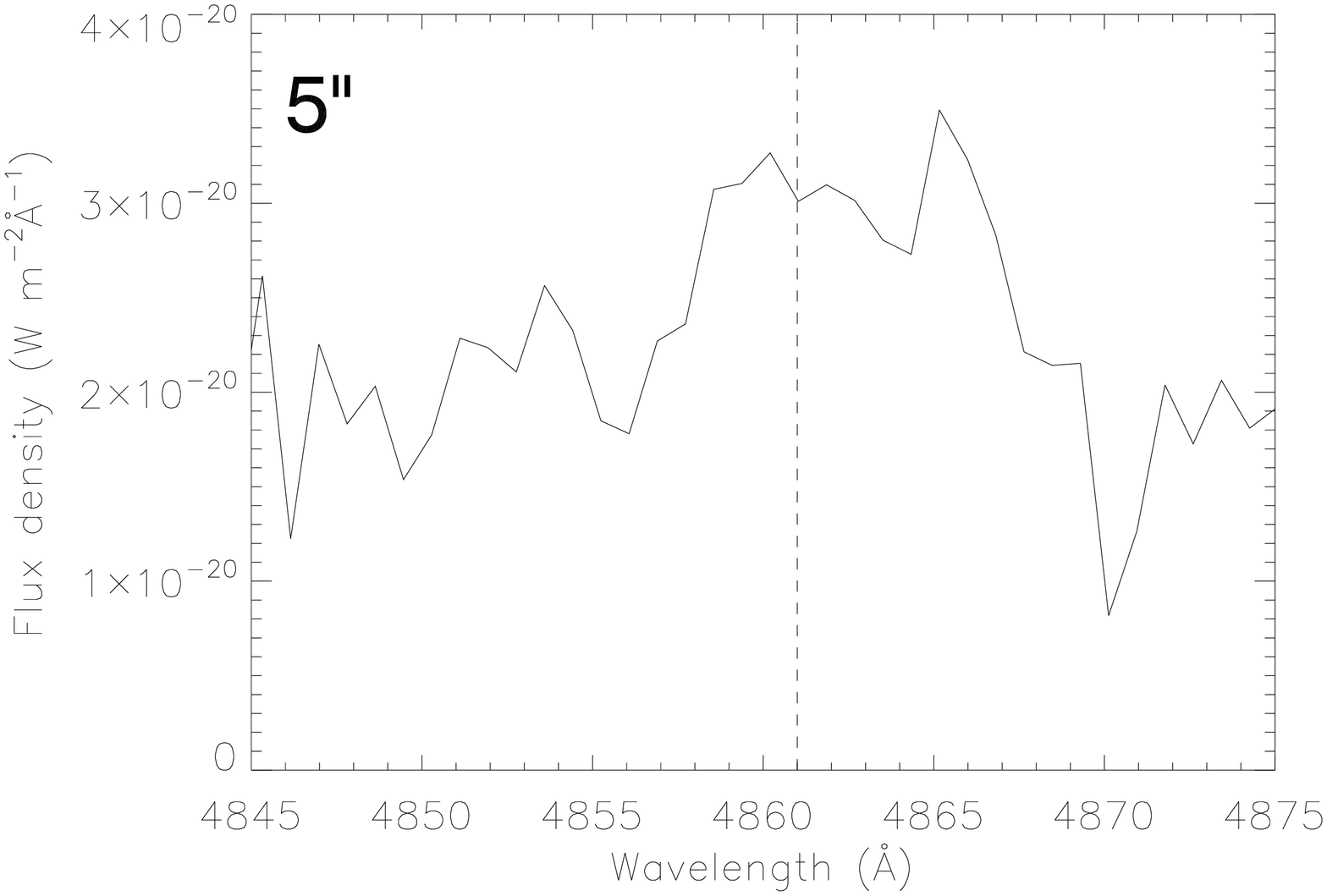}
\includegraphics[angle=90]{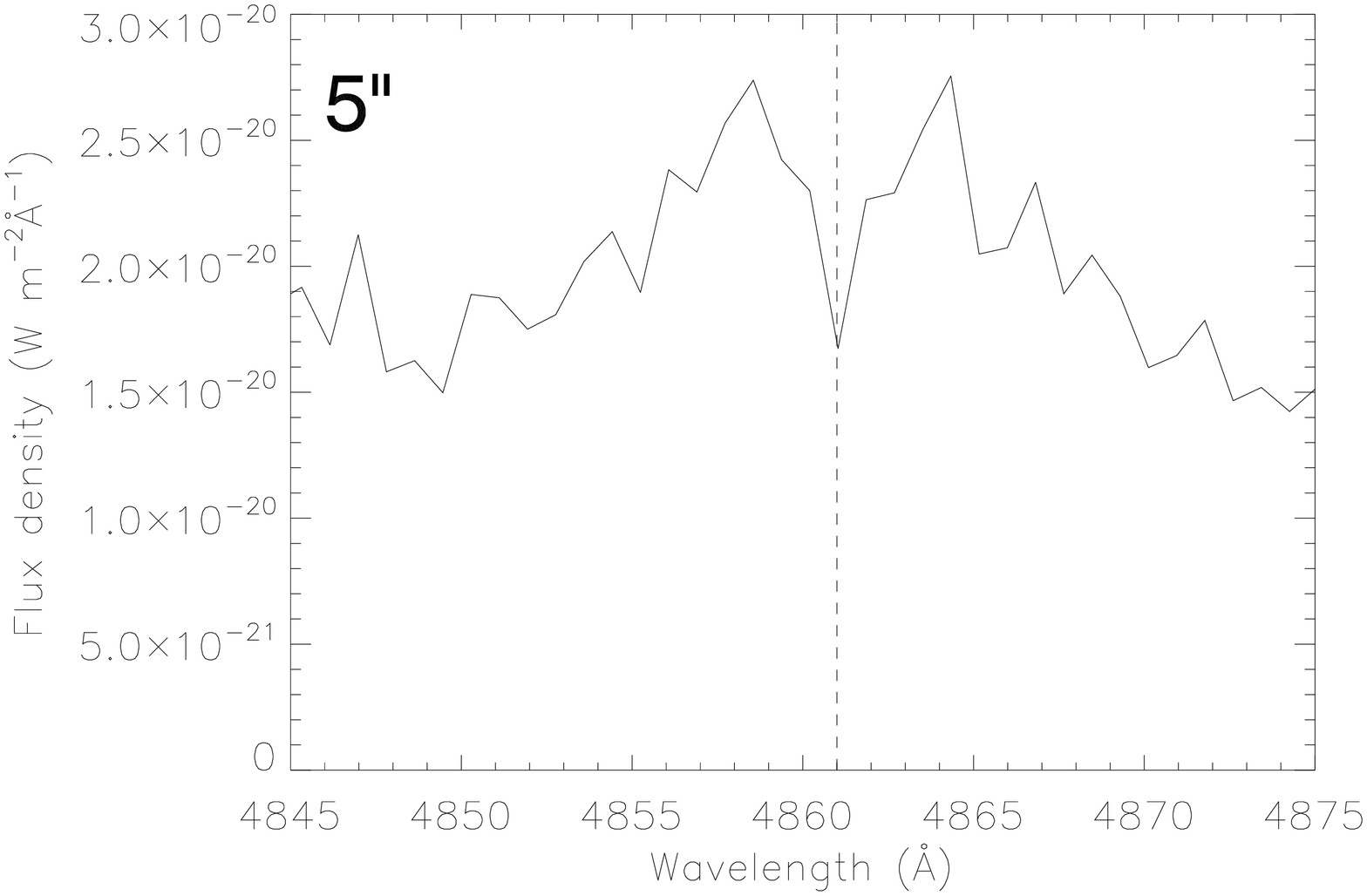}}\\
%\resizebox{\hsize}{!}
%{\includegraphics[angle=90]{b_7a_new.ps}
%\includegraphics[angle=90]{b_7b_new.ps}}\\
%\resizebox{\hsize}{!}
%{\includegraphics[angle=90]{b_10a_new.ps}
%\includegraphics[angle=90]{b_10b_new.ps}}

\caption{H$\beta$ line profiles for the parallel (left) and
  perpendicular (right) spectra, extracted using increasing aperture widths of (from
  top): 1\arcsec, 2\arcsec, 3\arcsec, 4\arcsec and 5\arcsec.  Balmer absorption is most clear in the perpendicular
  spectra, and generally becomes more pronounced with the inclusion of
  emission at greater distances
  from the AGN (i.e. the larger extraction apertures).}
\label{balmer}
\end{figure}

The material lying at larger angles to the radio source axis
(i.e. the regions probed by the perpendicular spectrum) is also of
interest. The line emission along this axis is far less extensive (as
expected for the regions lying on the edge of, or outside, the
ionization cone).  Its ionization state suggests a lower ionization
parameter $U$ (i.e. fewer and/or less energetic ionizing photons), and
the weak line emission in these regions could be better explained as
being due to ionization by young OB stars (HII regions).  Our continuum
modelling in the preceding section finds clear evidence for a young
stellar population with an age of roughly 5 million years, but does
not allow us to pinpoint its location.  However, the varying strength
of the Balmer absorption features in the parallel and perpendicular
spectra (Fig.~\ref{balmer}) can give us some idea of how widely
distributed this young stellar population might be.

This plot displays the H$\beta$ line profile for the parallel and
perpendicular spectra, using increasing extraction widths of 1\arcsec
to 5\arcsec, beyond which point increased aperture widths only lead to
a deterioration of the signal-to-noise.  In the perpendicular
extracted spectra, Balmer absorption can be clearly observed, and
becomes increasingly apparent as emission from regions at a greater
distance from the AGN is included (i.e. for increased aperture
widths).  The line emission in the parallel spectra is roughly
50-100\% more luminous (see Appendix A), and the evidence for
the same trend in the strength of the absorption component is less
clear-cut.  Whilst the line emission is strongest in the central
regions of the host galaxy, the distribution of the line absorption
must therefore be less well peaked, otherwise we would not observe a
strengthening of the absorption component in wider apertures.  This
suggests that any young stellar population
responsible for the Balmer absorption features is unlikely to lie
predominantly within the central few pc. The pronounced appearance of
the Balmer absorption in the perpendicular spectra suggests that the
underlying young stellar population is not linked to the location of
the growing radio source but rather, more widely distributed throughout the
galaxy. Indeed, given the relative ages of YSP and
radio source ($5 \times 10^6$years and $\sim 10^4$ years
respectively), a radio-source induced star formation scenario would
seem unlikely.

It is interesting to contrast the derived properties of the YSP 
detected in the spectra of 9C J1503+4528 with those found for other
sources.  Recent work on the GPS radio source PKS1345+12 (Rodriguez et
al, in prep) finds evidence for a very similar young stellar population to
that of 9C J1503+4528: an age of roughly 5 million years (c.f. $\sim
10^3$years for the radio source itself) and a mass of $\sim 10^6 
M_{\odot}$.  Imaging observations locate the YSPs in several super
star clusters lying in the outer 
regions of the host galaxy, rather than in a nuclear starburst ring
such as that observed for Cygnus A (e.g. Jackson, Tadhunter \& Sparks
1998).  
A further point of interest is the finding that the hosts of high
luminosity AGN observed as part of the Sloan Digital Sky Survey
(Kauffmann et al 2003) typically have younger mean stellar ages than
normal elliptical galaxies existing at the same epoch.  Additionally,
the young stars are not preferentially located near the galaxy
nuclei, but rather are spread out over a much wider scale.  These
properties are strikingly similar to those observed in 9C~J1503+4528.

\subsection{Gas kinematics}

The observed changes in gas kinematics and dominant ionization
mechanism in different regions of the EELR are not unexpected, given
the expected location of radio source shocks, but do provide two key
results: the clumpy ISM/IGM responsible for the EELRs observed around
powerful radio sources does appear to be in place prior to the expansion of
the radio source, and the kinematic properties of the extended regions
are relatively undisturbed.  Any disturbance of the ISM/IGM due to an
earlier interaction/merger has since settled down. 

\subsection{Spectroscopic evidence of a prior merger}

If the triggering of radio source activity in this galaxy was due to a
merger or interaction, the properties of any young stellar population
formed as part of that process can shed light on the exact mechanisms
and time scales involved.  Our continuum modelling of these spectra
confirms the presence of a young stellar population, whose formation
occurred several million years in advance of the onset of AGN
activity.  The mass of this YSP component is a low proportion of the
total stellar mass of the host galaxy. Together with the the quiescent
kinematics in the outer regions of the EELR, and the morphological
evidence described above, this suggests that {\it if} a  merger/interaction
was
responsible for both the eventual triggering of the radio source
activity and the formation of the young stellar population, it is most
likely to have been fairly minor in nature.

\section{Conclusions}

9C J1503+4528 provides a detailed insight into many of the events
associated with the triggering of a powerful radio source.  Our
observations have shown that:

\vspace{-0.2cm}
\begin{enumerate} 
\item[$\bullet$]The relative sizes of EELR and radio source (the
  former being $\approx 10$ times larger), together with typical
  compact radio source expansion velocities of 0.1 c, suggest
  that both the luminous quasar nucleus responsible for photoionizing
  the outer regions of the EELR, and the radio source itself, were
  triggered simultaneously.
\item[$\bullet$]The EELR ionization state is consistent with shocks
  being important on scales comparable to the radio source, with AGN
  photoionization within the ionization cones being the dominant
  mechanism further from the AGN, beyond the extent of the radio lobes.
\item[$\bullet$]The gas kinematics confirm that weak shocks could be
  important close to the radio source, and that the outer regions of
  the EELR are relatively undisturbed.  The kinematics suggest that
  motion along the radio axis may be taking place, which may indicate
  the presence of outflows.
\item[$\bullet$]The blue continuum emission and Balmer absorption
  features suggest that recent star formation has occurred, but their
  spatial distribution suggests that it is not caused by the passage
  of the radio source.
\item[$\bullet$]Modelling of the observed continuum emission gives a
  best-fit age for the young stellar population of $5 \times 10^6$
  years, and a mass of roughly 0.06\% of the total stellar mass of the
  host galaxy.
\item[$\bullet$]Imaging observations of the host galaxy reveal a
  simple elliptical morphology; fitting of the $K-$band image suggests
  an effective radius of $\sim 11.5$kpc.
\item[$\bullet$]A timescale of events can be built up from these
  results: star formation is triggered by some event; in this case,
  any tidal interaction or merger must have been relatively minor.
  AGN/radio source activity follows the onset of star formation by
  several million years.
\end{enumerate}
 
As a final consideration, we speculate on scenarios where the {\em
lack} of evidence for a recent merger may be significant. As noted in
the introduction, it is thought that the crucial ingredient for a
radio-loud AGN is that the central black hole is rapidly spinning, a
process which is most naturally explained by the coalescence of two
smaller black holes (Wilson \& Colbert 1995). Indeed, the cosmic
evolution of AGN activity can be well-described in terms of the
expected halo merger rate (Mahmood, Devriendt \& Silk 2005). However,
the timescales involved in the black hole coalescence after the halo
merger are extremely uncertain; while it is thought that SMBHs can be
brought within $\sim$ one parsec of each other on a timescale similar
to the dynamical friction timescale (e.g. Milosavljvic \& Merrit
2001), the mechanism for the final merger is unknown. 

One possibility
is that the formation of an accretion disc around one or both black 
holes provides the final necessary dynamical friction (e.g. Armitage
\& Natarajan 2005).   
If we assume that the central black hole of 9C J1503+4528 was spun up
as part of the 
AGN triggering process, the apparent
simultaneity of the triggering of both the optical AGN activity 
(requiring an accretion disc) and the radio jets (requiring the SMBH
to be spinning rapidly) for 9C J1503+4528  would demand that the final SMBH coalescence
must take place on a very short timescale indeed. If this were not the
case, and SMBH coalescence occurred more slowly, one would expect a
period of radio-quiet AGN activity to precede 
the formation of the jets, which is ruled out in the case of this
source by the relative sizes of 
the EELR and radio source.

Given the lack of evidence for a recent major merger, we therefore
propose that the host galaxy of 9C~J1503+4528 contained a rapidly 
spinning SMBH {\it prior} to  the present period of AGN activity. In this case,
the SMBH would have been spun up a significant time in the past, most
likely during one or more mergers of smaller gas-rich discs which, in
hierarchical clustering models, are the progenitors of
ellipticals (e.g. Springel et al 2005). This scenario has the advantage that radio-loud AGN may be
``re-lit'' by relatively small gas injections from time to time, which
provides a natural route for the necessary mechanical energy feedback
in the intracluster medium (Nipoti \& Binney 2005; Best et al 2005).  However, both
9C~J1503+4528 and the well-studied cluster radio sources are observed
at relatively recent cosmic time. The time delay between the initial
creation of a binary SMBH pair and the onset of jet activity may have
been significantly shorter in the early universe, where gas-rich
mergers were significantly more prevalent. Such a scenario could be
tested by investigating any time lag between the evolution of the
radio luminosity function at high redshift (which is still poorly
constrained, e.g. Willott et al. 2001) and the halo merger rate as
predicted by hierarchical clustering models.

\section*{Acknowledgements}

We thank Clive Tadhunter \& Joanna Holt for useful discussions and
allowing us to develop the CONFIT software for our own use; Daniel
Smith and Matt Jarvis for obtaining the $r$-band image; and Andy Adamson
for obtaining the $K$-band image.

Some of the data presented herein were obtained at the W. M. Keck
Observatory, which is operated as a scientific partnership among the
California Institute of Technology, the University of California, and
the National Aeronautics and Space Administration. The Observatory was
made possible by the generous financial support of the W. M. Keck
Foundation.

UKIRT is operated by the Joint Astronomy Centre on behalf of PPARC.

KJI acknowledges a PPARC research fellowship, DL a PPARC
PhD studentship and GC support from PPARC observational rolling grant
PPA/G/O/2003/00123.  

It is a pleasure to express our gratitude and
respect to the indigenous people of Hawai`i, from whose sacred
mountain Mauna Kea our observations were made.  

We thank the anonymous referee for their very useful comments on this paper.

\appendix
\section{Emission line properties of 9C\,J1503+4528}
\begin{table*}
\caption{Spectroscopic properties of 9C1503+4528, determined from
  various spectra obtained perpendicular and
  parallel to the radio source. The spectra were extracted with
  various width apertures, so as to probe the properties of the
  central and exterior regions of the extended emission structures.
  The total integrated fluxes for the \oo\ and H$\beta$ emission lines
  are provided, given in units of
  $10^{-17}\rm{W\,m}^{-2}$.  All other emission lines are quoted in
  terms of their flux ratios relative to both \oo\ and H$\beta$ within
  the extracted one-dimensional spectra.  This has been done due to
  the fact that the levels of Balmer absorption vary with position
  throughout the galaxy, and therefore a comparison with the H$\beta$
  line flux, though popular in the literature, may not be the most
  appropriate option for this source.  The errors on
  the
  H$\beta$ and [O\textsc{ii}] line fluxes are dominated by calibration errors, estimated to
  be $\la 10$\%. Errors on the other lines and this calibration error
  are added in quadrature with the errors arising from photon
  statistics. All quoted values
  in this table are corrected for galactic extinction using the
  $E(B-V)$ for the Milky Way (Schlegel, Finkbeiner \& Davis 1998) and the parametrized galactic extinction law of Howarth
  (1983). }
\begin{center} 
\begin{tabular}{lcccccc}
\multicolumn{2}{c}{Details of extracted spectrum} & Combined,
$2^{\prime\prime}$ & Parallel, $2^{\prime\prime}$ &
Perpendicular, $2^{\prime\prime}$ & Parallel, $5^{\prime\prime}$ &
Perpendicular, $5^{\prime\prime}$\\
\multicolumn{2}{c}{\oo\ flux} & $191.5$& $227.5$&$158.6$ &$248.0$ & $180.2$\\
\multicolumn{2}{c}{H$\beta$ flux} & $11.7$&$13.4$&$10.0$ &$14.4$ &$10.2$ \\\\
%\hline
{\oo 3727}            & \oo\ Flux Ratio      & 100        & 100        &100         &100         & 100 \\
                      & H$\beta$ Flux Ratio &$1635\pm177$&$1701\pm191$&$1588\pm195$&$1715\pm231$&$1766\pm290$\\
                      & Equiv. width        & 182\AA     & 192\AA     & 170\AA     & 150\AA     & 123\AA\\
\ne3 3869             & \oo\ Flux Ratio      & $8.5\pm0.9$& $8.7\pm1.0$&$8.1\pm1.0$ &$9.5\pm1.1$ &$11.1\pm1.5$ \\
                      & H$\beta$ Flux Ratio &$139\pm16.2$&$148\pm18.2$&$128\pm17.9$&$162\pm24.3$&$196\pm36.7$ \\
                      & Equiv. width        & 17\AA      & 19\AA      & 15\AA      & 16\AA      & 15\AA\\
H$\zeta$ 3889         & \oo\ Flux Ratio      & $0.7\pm0.2$& $1.6\pm0.3$&$1.1\pm0.3$ &$1.3\pm0.5$ &$<0.3$ \\
                      & H$\beta$ Flux Ratio &$12.1\pm3.4$&$26.4\pm5.8$&$16.7\pm5.6$&$21.6\pm7.9$&$<5.2$ \\
                      & Equiv. width        & 1.3\AA     & 3.1\AA     & 1.9\AA     & 1.9\AA     & $<$0.2\AA\\
H$\epsilon$+\ne3 3967 & \oo\ Flux Ratio      & $0.6\pm0.2$& $1.0\pm0.2$&$<0.3$ &$0.6\pm0.3$ &$<0.6$ \\
                      & H$\beta$ Flux Ratio & $9.1\pm3.1$&$16.4\pm3.0$&$<5.3$ &$10.8\pm4.8$&$<10.3$ \\
                      & Equiv. width        & 1.0\AA     & 1.9\AA     & $<$0.3\AA     & 1.0\AA     & $<$0.4\AA\\
\s2  4072             & \oo\ Flux Ratio      & $2.7\pm0.4$& $3.0\pm0.4$&$2.7\pm0.5$ &$1.9\pm0.4$ &$2.0\pm0.7$ \\
                      & H$\beta$ Flux Ratio &$43.7\pm6.5$&$50.0\pm7.3$&$43.3\pm9.1$&$31.7\pm7.0$& $35.0\pm13.5$\\
                      & Equiv. width        & 4.1\AA     & 4.9\AA     & 3.9\AA     & 2.3\AA     & 2.1\AA\\
H$\delta$ 4102        & \oo\ Flux Ratio      & $0.7\pm0.3$& $0.9\pm0.4$&$0.6\pm0.4$ &$0.8\pm0.7$ &$<0.5$ \\
                      & H$\beta$ Flux Ratio &$11.6\pm4.8$&$14.4\pm6.3$&$9.4\pm6.0$ &$13.6\pm12.4$&$<8.3$ \\
                      & Equiv. width        & 1.1\AA     & 1.5\AA     &0.9\AA      &1.1\AA      & $<$0.4\AA\\
H$\gamma$  4340       & \oo\ Flux Ratio      & $3.3\pm0.5$& $3.7\pm0.5$&$2.8\pm0.5$ &$3.7\pm0.6$ &$1.5\pm0.5$ \\
                      & H$\beta$ Flux Ratio &$54.2\pm7.7$&$63.0\pm8.7$&$44.1\pm8.5$&$62.2\pm11.3$&$25.4\pm10.1$ \\
                      & Equiv. width        & 5.7\AA     & 6.9\AA     & 4.3\AA     &5.3\AA      & 1.7\AA\\
\o3 4363              & \oo\ Flux Ratio      & $1.2\pm0.3$& $1.1\pm0.3$&$0.8\pm0.3$ &$1.4\pm0.4$ &$0.7\pm0.4$ \\
                      & H$\beta$ Flux Ratio &$19.8\pm5.1$&$19.1\pm5.6$&$12.8\pm4.8$&$23.2\pm7.4$&$12.0\pm6.7$ \\
                      & Equiv. width        & 2.1\AA     & 2.0\AA     & 1.2\AA     & 1.9\AA     & 0.8\AA\\
H$\beta$  4861        & \oo\ Flux Ratio      & $6.1\pm0.7$& $5.9\pm0.7$&$6.3\pm0.8$ &$5.8\pm0.8$ & $5.7\pm0.9$\\
                      & H$\beta$ Flux Ratio & 100        & 100        & 100        & 100        & 100\\
                      & Equiv. width        & 10\AA      & 10\AA      &8.9\AA      & 7.7\AA     & 6.0\AA\\
\o3 4959              & \oo\ Flux Ratio      &$10.7\pm1.1$&$9.9\pm1.1$&$10.5\pm1.2$&$13.5\pm1.6$&$12.4\pm1.6$ \\
                      & H$\beta$ Flux Ratio &$175\pm19.7$&$169\pm20.2$&$167\pm21.8$&$232\pm33.5$&$219\pm39.6$ \\
                      & Equiv. width        & 17\AA      & 17\AA      & 16\AA      &18\AA       & 14\AA\\
\o3 5007              & \oo\ Flux Ratio      &$22.6\pm2.4$&$23.7\pm2.5$&$24.1\pm2.7$&$30.5\pm3.2$&$25.2\pm3.3$ \\
                      & H$\beta$ Flux Ratio &$370\pm41.2$&$403\pm46.3$&$383\pm50.7$&$523\pm72.9$&$446\pm83.1$ \\
                      & Equiv. width        & 38\AA      & 43\AA      & 40\AA      & 41.8\AA    & 32\AA\\
\end{tabular}                      
\end{center}                       
\end{table*}

\begin{table}
\caption{Changes to the best-fit YSP model with the inclusion of a power law
  component.  Columns 1 and 2 give the percentage of the power-law component
  at a wavelength of $\sim 3600$\AA, and its spectral index, $\alpha$
  (defined as $F_\lambda \propto \lambda^\alpha$). Column 3 gives the
  best-fit old stellar population age in Gyr.  The best-fit young
  stellar population age (in Myr) and mass (as a percentage of the
  total stellar mass of the host galaxy) are given in columns 4 \& 5
  respectively.  The reduced $\chi^2$ for the best-fit model is listed
  in column 6.}
\begin{center} 
\begin{tabular}{cccccc}
\footnotesize
{Power law \% } & {$\alpha$} & {Old SP} &
\multicolumn{2}{c}{YSP properties}& 
{Reduced}\\
{at $\sim 3600$\AA}& { }& { age  (Gyr)}& {age (Myr)}&{mass \%}& $\chi^2$ \\
(1) & (2) & (3) & (4) & (5) & (6)\\ \hline
0.0\%     &  N/A   &  6.0  & 5  &  0.065\% & 1.01943  \\%73.5\% & 
1.0\%     &  -1  &  6.0 & 5 &  0.064\% & 1.03379  \\%73.4\% & 
1.0\%     &  -2  &  6.0 & 5 &  0.064\% & 1.03269  \\%73.3\% & 
1.0\%     &  -3  &  6.0 & 5 &  0.064\% & 1.03236  \\%73.3\% & 
2.5\%     &  -1  &  6.0 & 5 &  0.063\% & 1.05603  \\%73.2\% & 
2.5\%     &  -2  &  6.0 & 5 &  0.063\% & 1.05338  \\%73.1\% & 
2.5\%     &  -3  &  6.0 & 5 &  0.062\% & 1.05281  \\%73.0\% & 
5.0\%     &  -1  &  6.0 & 5 &  0.061\% & 1.09503  \\%72.9\% & 
5.0\%     &  -2  &  6.0 & 5 &  0.060\% & 1.09005  \\%72.6\% & 
5.0\%     &  -3  &  6.0 & 5 &  0.060\% & 1.08985  \\%72.4\% & 
10.0\%    &  -1  &  6.0 & 5 &  0.058\% & 1.18089  \\%72.2\% & 
10.0\%    &  -2  &  5.0 & 5 &  0.063\% & 1.17021  \\%69.5\% & 
10.0\%    &  -3  &  5.0 & 5 &  0.061\% & 1.17095  \\%69.0\% & 
15.0\%    &  -1  &  5.0 & 5 &  0.061\% & 1.27639  \\%69.2\% & 
15.0\%    &  -2  &  5.0 & 5 &  0.058\% & 1.26151  \\%68.1\% & 
15.0\%    &  -3  &  5.0 & 5 &  0.056\% & 1.26901  \\%67.4\% & 
25.0\%    &  -1  &  5.0 & 5 &  0.052\% & 1.50695  \\%67.0\% & 
25.0\%    &  -2  &  5.0 & 5 &  0.048\% & 1.49007  \\%64.8\% & 
25.0\%    &  -3  &  7.0 & 50 &  0.703\% & 1.50599  \\%68.9\% & 
50.0\%    &  -1  &  6.0 & 50 &  0.489\% & 2.20266  \\%62.4\% & 
50.0\%    &  -2  &  6.0 & 100 &  0.774\% & 1.96529  \\%53.0\% & 
50.0\%    &  -3  &  7.0 & 100 &  0.647\% & 1.83815  \\%53.8\% & 
\end{tabular}
\end{center}
\end{table}

Here we present the results of our emission-line fitting for the  parallel,
perpendicular and combined 2-d frames. 
Fluxes relative to both \oo 3727\AA\ and $H\beta$ are tabulated in
Table A1, as well as their equivalent widths. Although flux ratios
quoted relative to $H\beta$ are more common, the presence of Balmer
absorption lines reduces the usefulness of this ratio.  The values
in Table A1 have been corrected for galactic extinction, using values
for the H\textsc{i} column density of the Milky Way taken from the
NASA Extragalactic Database (NED), and the parametrized galactic
extinction law of Howarth (1983).

\section{Best-fit YSP parameters in the presence of a power-law.}
As part of our modelling, we consider the effects of the inclusion of
a power-law component on the best-fit YSP parameters.  The preferred
age of the young stellar population remains unchanged at 5\,Myr except for the
strongest power-law contributions ($> 25\%$).  However, the reduced
$\chi^2$ values indicate that models with a strong power law
contribution are not favoured.
\label{lastpage}

\end{document}